\documentclass[10pt,preprint]{aastex}

\usepackage{amsmath, amsthm, amssymb}
\usepackage{natbib}
\usepackage{epsfig}
\usepackage[update,prepend]{epstopdf}

\shorttitle{Four new white dwarf companions in {\it  Kepler} }
\shortauthors{Faigler et al.}

\begin{document}
\title{BEER analysis of {\it Kepler} and {\it CoRoT} light curves: \\
IV. Discovery of four new low-mass white dwarf companions in the {\it Kepler} data}

\author{S. Faigler\altaffilmark{1},  I. Kull\altaffilmark{1}, T. Mazeh\altaffilmark{1}, F. Kiefer\altaffilmark{1},
           D. W. Latham\altaffilmark{2} and S. Bloemen\altaffilmark{3}}

\altaffiltext{1}{ School of Physics and Astronomy, Raymond and Beverly Sackler Faculty of Exact Sciences, Tel Aviv University, Tel Aviv  69978, Israel}

\altaffiltext{2}{ Harvard-Smithsonian Center for Astrophysics, 60 Garden St., Cambridge, MA 02138, USA}
\altaffiltext{3}{ Department of Astrophysics, IMAPP, Radboud University Nijmegen, 
P.O. BOX 9010, NL-6500 GL Nijmegen, the Netherlands}

\begin{abstract}
We report the discovery of four short-period eclipsing systems in the {\it Kepler} light curves, consisting of an A-star primary and a low-mass white dwarf (WD) secondary (dA+WD)---KIC 4169521, KOI-3818, KIC 2851474, and KIC 9285587. 
The systems show BEaming, Ellipsoidal and Reflection (BEER) phase modulations together with primary and secondary eclipses.
These add to the 6 {\it Kepler} and 18 WASP short-period eclipsing dA+WD binaries that were previously known.
The light curves, together with follow-up spectroscopic observations, allow us to derive the masses, radii, and effective temperatures of the two components of the four systems. 
The orbital periods, of $1.17$--$3.82$ days, and WD masses, of $0.19$--$0.22$ $M_{\odot}$, are similar to those of the previously known systems. 
The WD radii of KOI-3818, KIC 2851474, and KIC 9285587 are $0.026$, $0.035$, and $0.026$ $R_{\odot}$, respectively, the smallest WD radii derived so far for short-period eclipsing dA+WD binaries.
These  three binaries extend the previously known population to older systems with cooler and smaller WD secondaries.
KOI-3818 displays evidence for a fast-rotating primary and a minute but significant eccentricity, $\sim 1.5\times10^{-3}$. These features are probably the outcome of the mass-transfer process.
\end{abstract}


\section{Introduction}
The {\it Kepler} spacecraft was launched in order to detect shallow transits produced by planets, which are characterized by their small radii, of the order of $0.1$--$0.01$ $R_{\odot}$, and therefore induce transits with shallow depth, of the order of $10^{-2}$--$10^{-4}$ of the stellar flux \citep{borucki10}. As of 2015 May, {\it Kepler} indeed produced more than $4600$ planet candidates, with orbital periods of $0.3$--$1295$ days \citep{mullally15}. As a by-product of this effort, the {\it Kepler} mission has also identified more than 2700\footnote[4]{http://keplerebs.villanova.edu/} eclipsing binary (EB) systems  \citep{slawson11}, most of which exhibit much deeper eclipses.  

However, white dwarfs (WDs) residing in binary systems are also expected to produce shallow eclipses, mimicking the transits of small planets.
In fact, there are nine known EBs with WD secondaries in the {\it Kepler} data \citep{rowe10,bloemen11,carter11,breton12,muirhead13,kruse14,rappaport15}. Six of these were identified as short-period binaries of an F/A-type primary and a pre-Helium WD secondary (dA+pre-He-WD), with radii of $0.04$--$0.28$ $R_{\odot}$. The remaining three systems belong to different categories and had more ``standard'' derived WD radii of $0.012$--$0.014$ $R_{\odot}$. 

Other then the {\it Kepler} discoveries, the stars V209, 1SWASP J024743.37-251549.2 (hereafter WASP0247-25), OGLE-BLG-RRLYR-02792, and possibly AW UMa, are EBs that are believed to accommodate pre-He-WD secondaries \citep{kaluzny07,pribulla08,maxted11,maxted13,pietrzynski12}.
Of these, only WASP0247-25 was identified as a short-period detached dA+pre-He-WD EB, which we consider as the binary type associated with the discoveries reported here.
A significant recent contribution to the known population of short-period dA+pre-He-WD EBs was made by \citet{maxted14}, who discovered 17 such systems in the WASP photometry database \citep{pollacco06}. Two of these have accurate measurements of the secondaries' masses, radii, and $T_{\rm eff}$, confirming that they are pre-He-WDs.

WDs in short-period binaries are expected to be hotter than their primary stars, resulting in a light curve with a flat-bottom secondary eclipse that is deeper than the primary eclipse. This is because for a circular orbit, the primary-to-secondary eclipses depth ratio approximates the primary-to-secondary surface-brightness ratio in the observed band. 
 Indeed, except for the special case of KPD 1946+4340, with a subdwarf B star (sdB) primary \citep{bloemen11}, the other eight WD systems in the {\it Kepler} data exhibit 
 a secondary eclipse that is deeper than the primary one.

Unfortunately, just from the primary and secondary eclipse depths, one cannot tell if the companion is a low-mass star or a WD, as the primary and secondary eclipses can be interchanged. 
To overcome this ambiguity, \citet{maxted14} looked in the WASP catalog for EB systems that show a {\it deeper flat-bottom} eclipse, indicating a hotter WD companion. Using this method they discovered 17 short-period EBs with pre-He-WD secondaries. More recently, \citet{rappaport15} used a similar approach for the {\it Kepler} EB catalog. They visually inspected the light curves of EBs with {\it Kepler} Input Catalog (KIC) $T_{\rm eff}>7000$ K and discovered two new short-period dA+pre-He-WD EBs .

Alternatively, one can use three photometric phase modulation effects, BEaming, Ellipsoidal and Reflection (BEER), to distinguish between low-mass stellar and WD companions. The beaming effect, sometimes called Doppler boosting, causes an increase (decrease) of the brightness of any light source approaching (receding from) the observer \citep{rybicki79,loeb03}, with an amplitude that is proportional to the radial velocity (RV) of the source. Therefore, the stellar RV modulation due to a circular-orbit companion will produce a sine-like beaming modulation at the orbital period, if the middle of the primary eclipse is defined as the phase zero point. The semi-amplitude of such a modulation is on the order of $100$--$400$ parts-per-million (ppm) for low stellar-mass companions, compared to an order of $10$ ppm for Jupiter-mass planets \citep{loeb03,faigler12}. 
More importantly, the phase of the beaming effect reveals which object is being eclipsed, and thus it enables distinguishing between a low-mass stellar companion and a WD companion, based on the eclipses' relative depth. 
 It is the same as the information that is provided by the phases when the radial velocity of the primary is blue shifted or red shifted.

The second effect is the well-known ellipsoidal variation \citep{kopal59,morris85} that is due to the tidal distortion of the star by the gravity of the companion \citep[e.g.,][]{loeb03,zucker07,mazeh08}, resulting in a cosine-like phase modulation at half the orbital period, for a circular-orbit companion under the same phase zero definition. The semi-amplitude of the ellipsoidal modulation for orbital periods of a few days is on the order of $1000$ ppm for low stellar-mass companions, compared to an order of $10$ ppm for Jupiter-mass planets \citep{loeb03,faigler12}.

The third effect is the reflection/emission variation, the result of light emitted by one component, scattered off of or thermally re-emitted from the dayside of the other component \citep{vaz85,wilson90,maxted02,harrison03,for10,reed10}. 
This effect depends on properties that are associated with the response of the object's atmosphere to its companion's radiation, such as the Bond albedo, the scattered light geometric albedo, and heat redistribution parameters. 
The reflection/emission modulation is expected to behave approximately as a cosine wave at the orbital period for a circular orbit, and can have different signs depending on the luminosity ratio and radius ratio of the two components of the binary. In the known {\it Kepler} systems with large-radii WD secondaries, this effect is dominated by light originating from the WD, scattered off of or thermally re-emitted by the primary star atmosphere \citep{carter11,breton12,rappaport15}.

To take advantage of the information provided by the BEER modulations, the BEER algorithm \citep{faigler11} searches for stars whose light curves show a combination of the three effects'  amplitudes and phases that is consistent with a short-period companion.
This work reports on the discovery of four additional short-period EBs of an A-type primary and a low-mass WD secondary  (dA+WD) in the {\it Kepler} field, 
identified among the BEER compact companion candidates as having a secondary eclipse deeper than the primary one.
One system includes a pre-Helium WD, similar to those in the previously known systems, while the other secondaries are well-developed WDs, which were not observed before in short-period dA+WD binaries.

This paper is organized as follows.
Section~\ref{sec:beer} presents the BEER search and the resulting four detections, and
Section~\ref{sec:spec} describes the spectroscopic follow-up observations. 
Section~\ref{sec:model} then presents detailed modeling of the {\it Kepler} photometry, and Section~\ref{sec:params} describes the parameters resulting from the analyses of the photometric and spectroscopic observations.
Section~\ref{sec:indiv} further reviews specific features of each of the four discovered systems, and
Section~\ref{sec:disc} summarizes and discusses the results of this work. 

\section{The Photometric BEER search}
\label{sec:beer}
To identify WD secondaries, we applied the BEER search algorithm, after adaptation for a compact object companion, to the {\it Kepler} Q2--Q16 raw long-cadence light curves of $40,728$ stars that were brighter than $13.5$ mag. 
 The BEER search assigned to each light curve a likelihood that the star hosts a compact companion, while identifying the inferior conjunction from the amplitudes and phases of the BEER modulations. After sorting the stars based on their BEER likelihoods, we visually inspected the $100$ highest-scoring light curves and identified four systems, KIC 4169521, KOI-3818=KIC 6515722, KIC 2851474, and KIC 9285587, in which there were two eclipses and the secondary was deeper than the primary.
As indicated by its name, KOI-3818 was a member of the KOI catalog, listed as a false positive, but with an orbital period of $\sim$$1.9$ days that is half the orbital period we detected, while the remaining three systems appeared in the {\it Kepler} EB catalog. 
 
\begin{figure*} 
\centering
\resizebox{18cm}{4.8cm}
{
\includegraphics{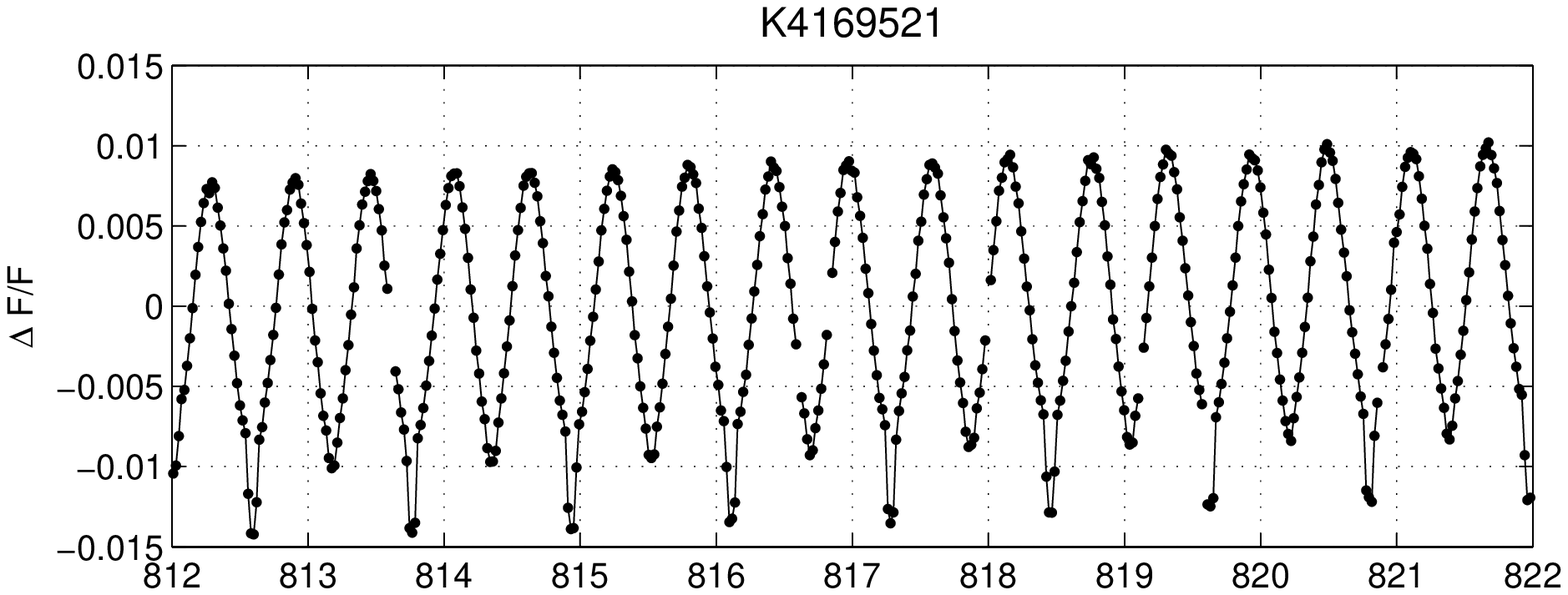}
}
\centering
\resizebox{18cm}{4.8cm} 
{
\includegraphics{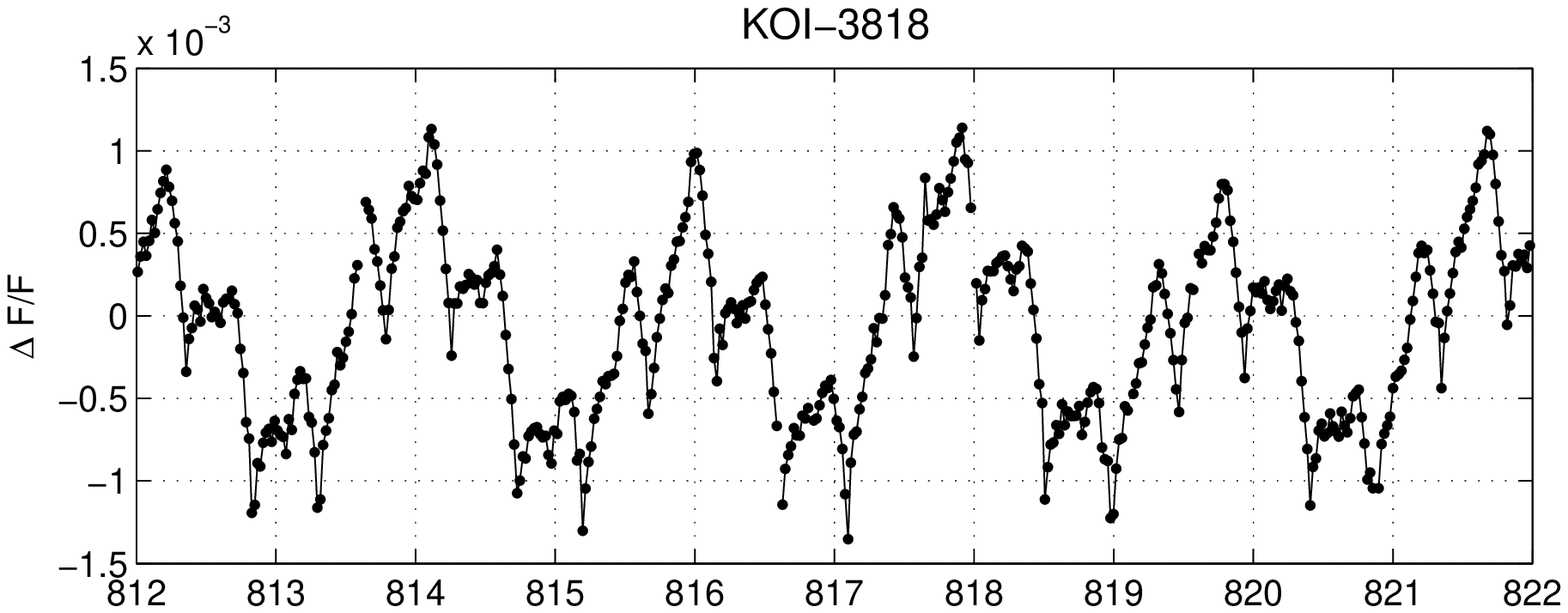}  
}
\centering
\resizebox{18cm}{4.8cm} 
{
\includegraphics{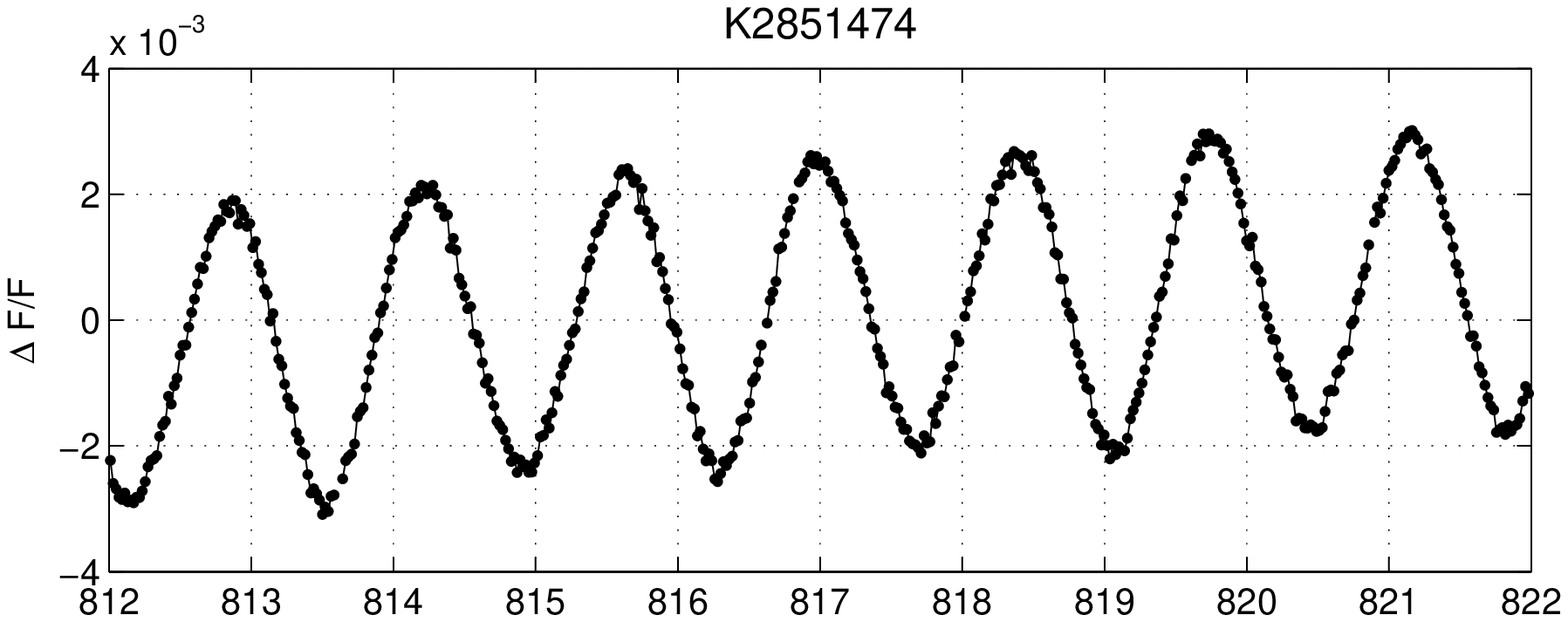}  
}
\centering
\resizebox{18cm}{4.8cm} 
{
\includegraphics{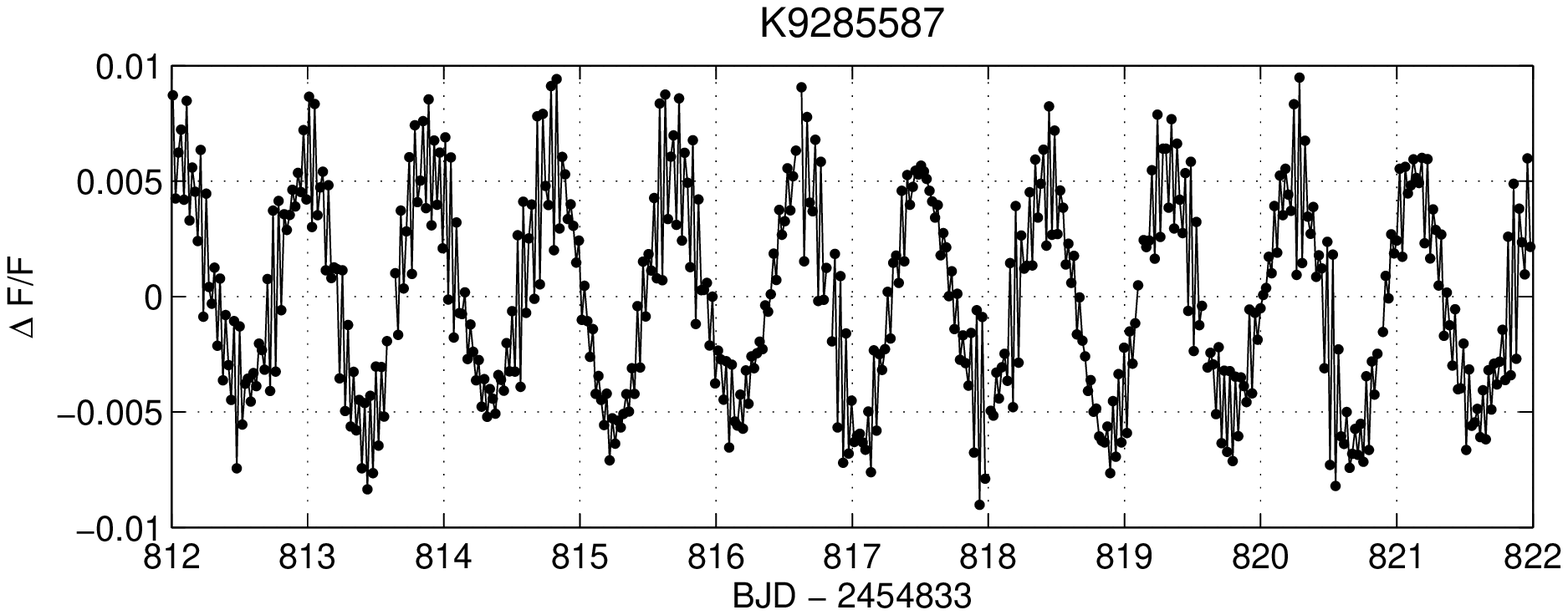}  
}

\caption{
Light curves of the four systems for a selected time span of 10 days. 
Note the different Y axis scales of the four plots. 
}
\label{fig:lc}
\end{figure*}

\begin{figure*} 
\centering
\resizebox{18cm}{4.5cm}
{
\includegraphics{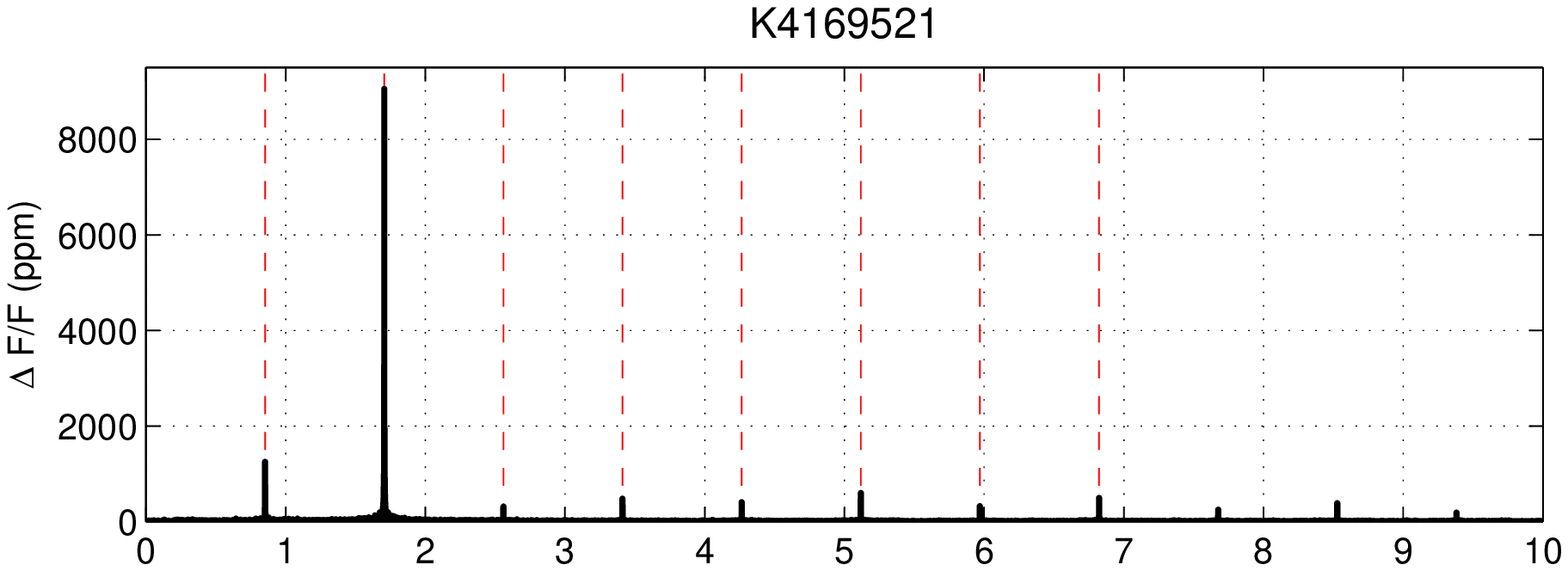}
}
\centering
\resizebox{18cm}{4.5cm} 
{
\includegraphics{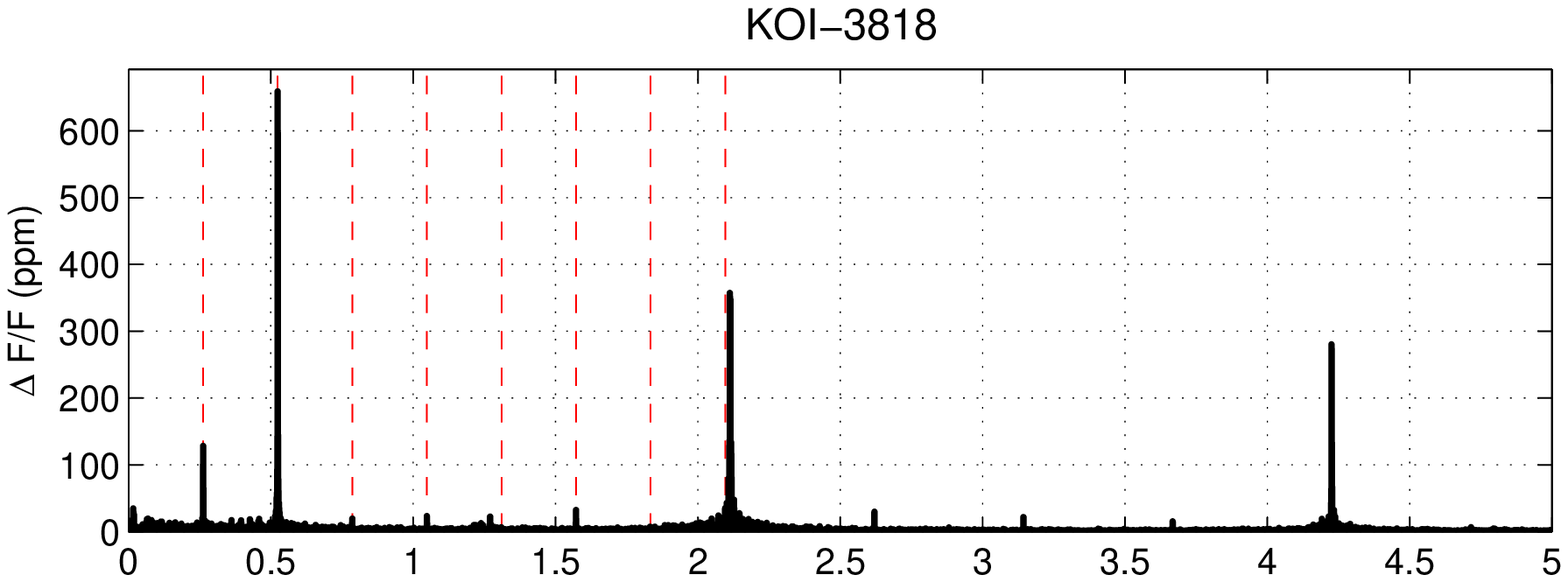}  
}
\centering
\resizebox{18cm}{4.5cm} 
{
\includegraphics{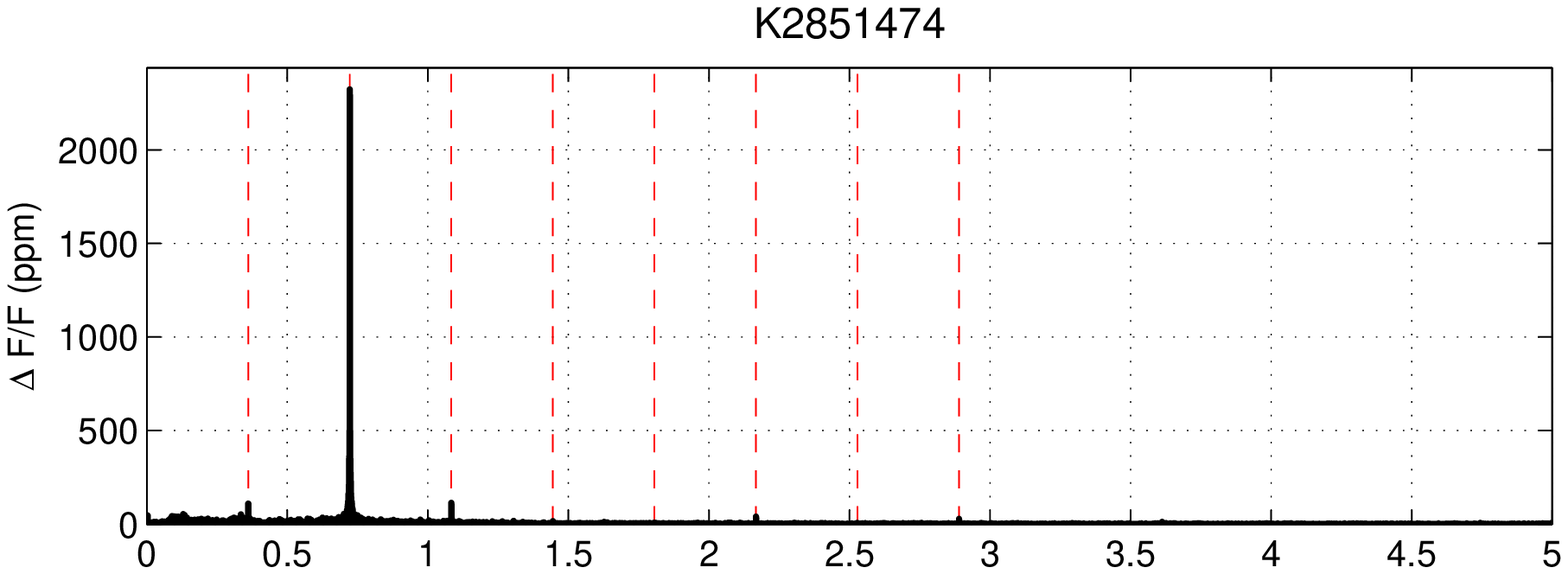}  
}
\centering
\resizebox{18cm}{4.5cm} 
{
\includegraphics{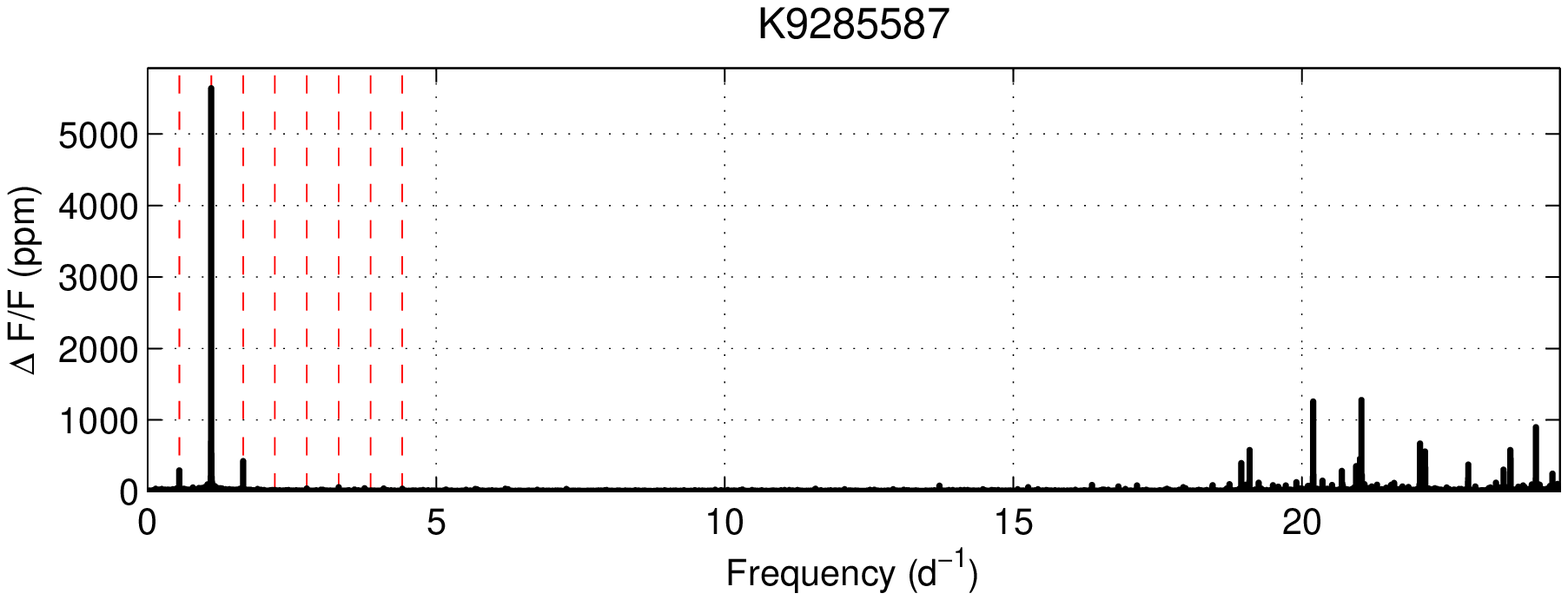}  
}
\caption{
Amplitude spectra of the four systems. Dashed lines mark the orbital frequency and its harmonics, determined from the eclipses' timings.
The frequency axis range is different for each system for improved visualization of specific spectral features of that system.
}
\label{fig:psd}
\end{figure*}

\begin{figure*} 
\centering
\resizebox{17cm}{9.0cm}
{
\includegraphics{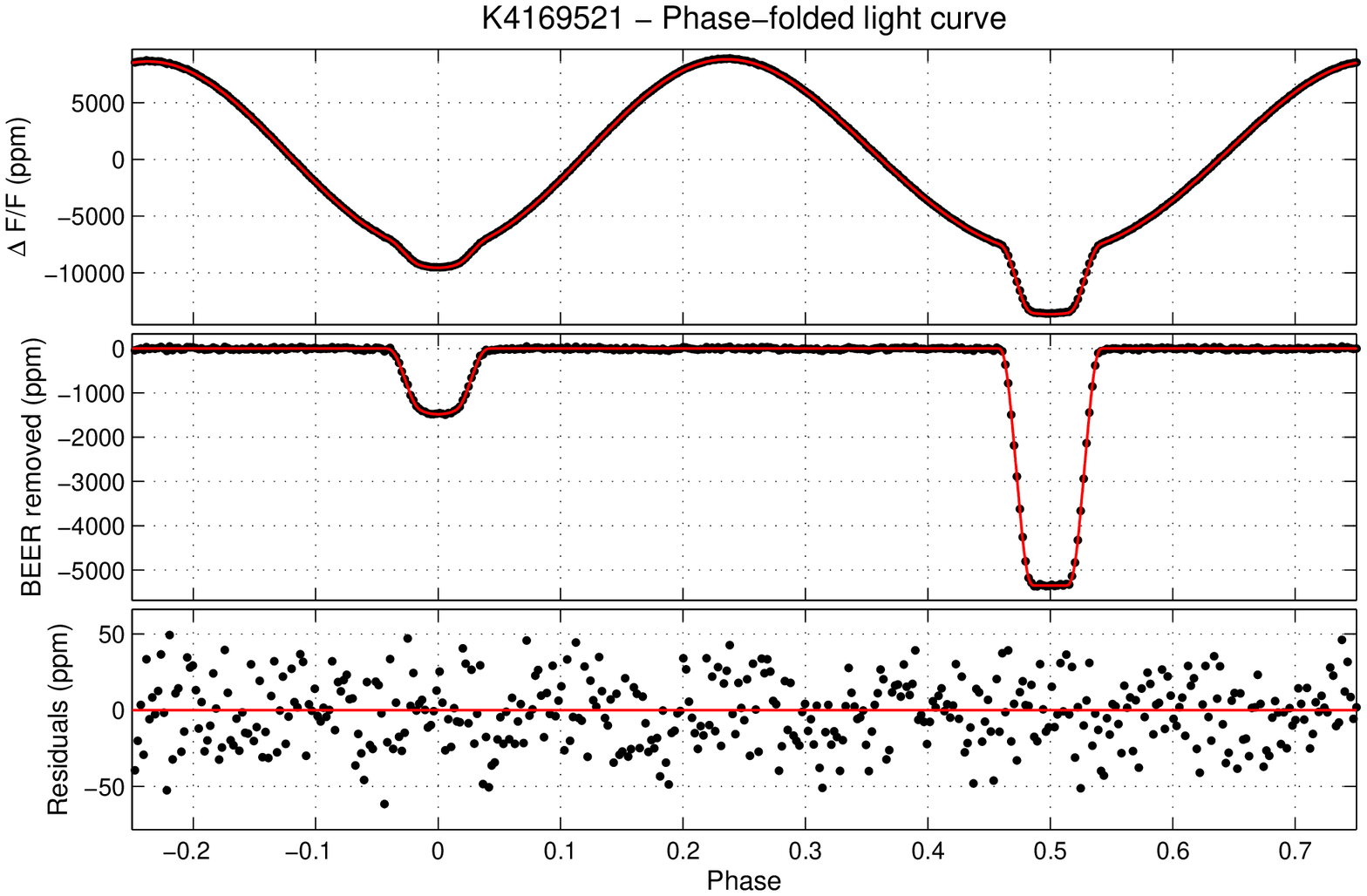}  
}
\centering
\resizebox{17cm}{9.0cm}
{
\includegraphics{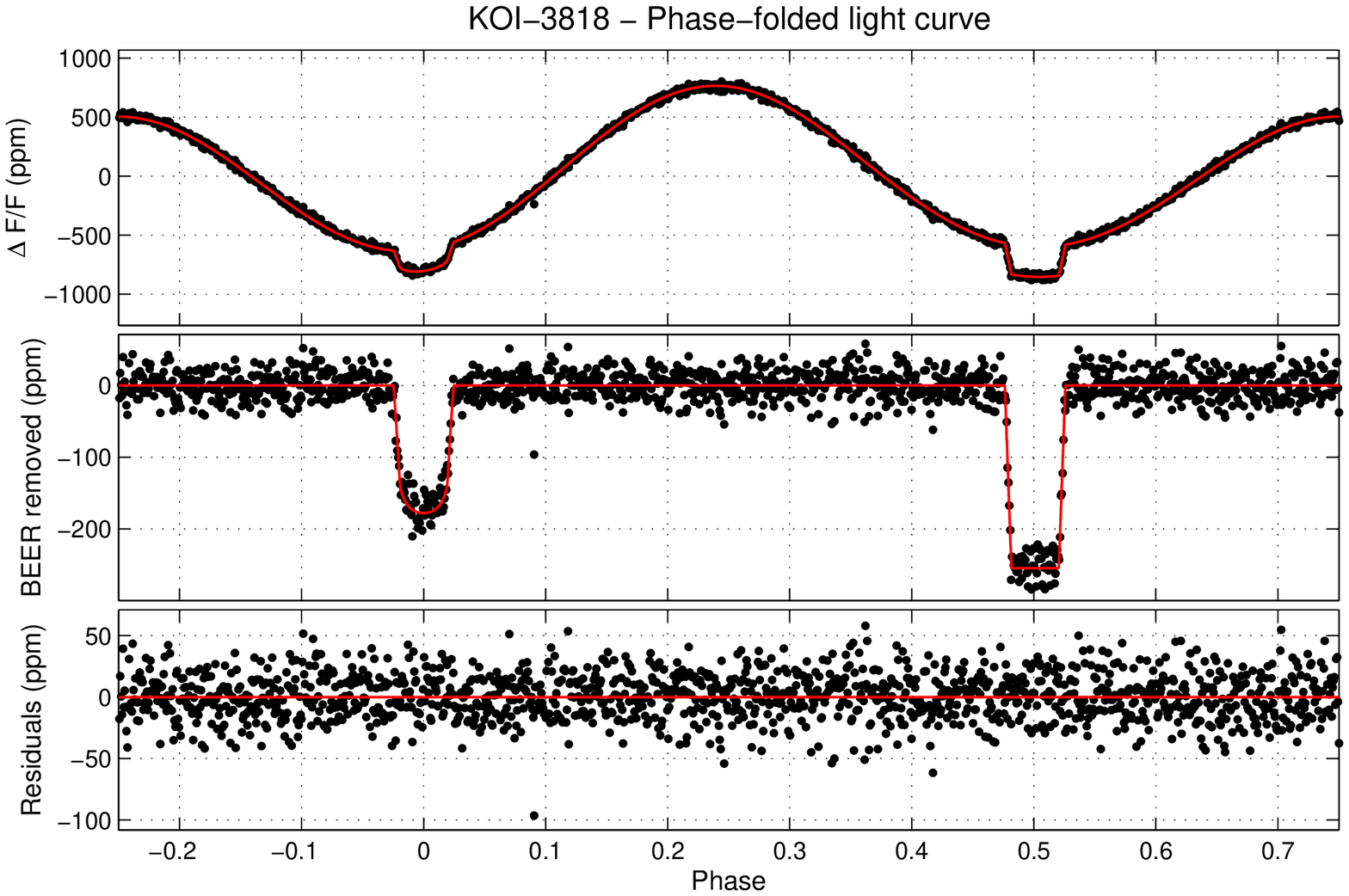}  
}
\caption{
The folded cleaned and detrended light curves of the four systems. The upper panels present the folded data, the middle panels show the folded data after the removal of the BEER modulation, and the lower panels present the residuals. The black points are the folded data in $4$ minutes bins, and the red curve is the model fit described in Section~\ref{sec:model}.
}
\label{fig:fold}
\end{figure*}

 Figure~\ref{fig:lc}  presents a short section of the raw {\it Kepler} light curve of each of the systems, Figure~\ref{fig:psd} shows the amplitude spectrum of the 
light curves, and Figure~\ref{fig:fold} presents the four cleaned and detrended light curves folded at their respective orbital periods. Cleaning of outliers and jumps and detrending were performed following \citet{faigler13}.
In the four systems the amplitude spectrum clearly shows the BEER frequency peaks at the orbital frequency and its first harmonic, associated mainly with the beaming and the ellipsoidal modulations, respectively. The KOI-3818 spectrum also shows peaks at frequencies of $\sim$$2.1$ cycles day$^{-1}$ and its harmonics; KIC 9285587 displays peaks in the $19$--$24$ cycles day$^{-1}$ range. Therefore in these two cases, cleaning also included fitting and subtracting several high-frequency sine functions, associated with the high-frequency spectra peaks (see Section~\ref{sec:model} for more details about the cleaning and detrending process).
 Each of the resulting folded light curves in Figure~\ref{fig:fold} shows ellipsoidal and beaming phase modulation, a curved bottom primary eclipse
 at phase zero, and a deeper flat-bottom secondary eclipse at phase $0.5$, all being consistent with a fully occulted, compact hot companion.

\begin{figure*} 
\centering
\resizebox{17cm}{9.0cm}
{
\includegraphics{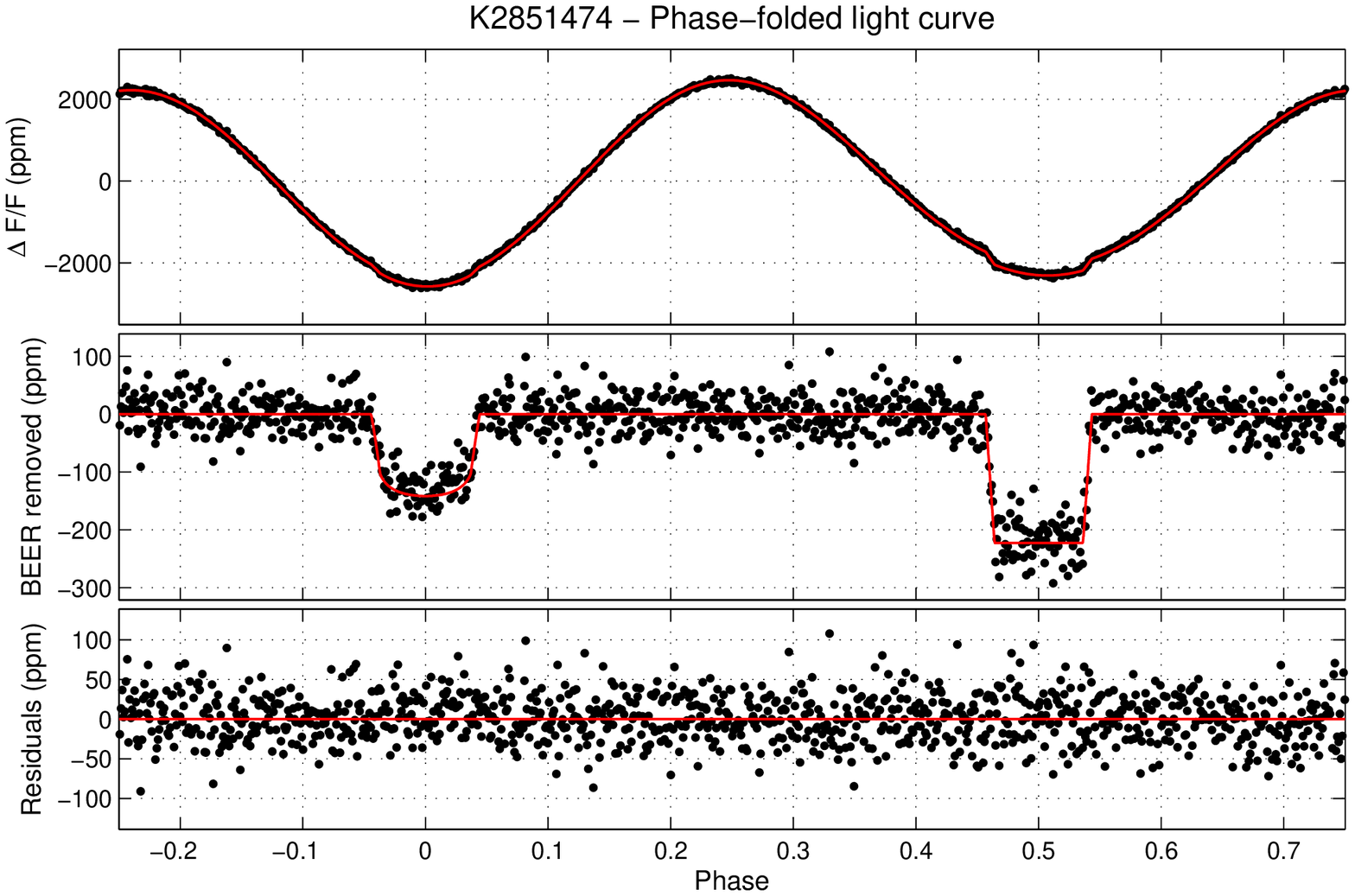}  
}
\centering
\resizebox{17cm}{9.0cm}
{
\includegraphics{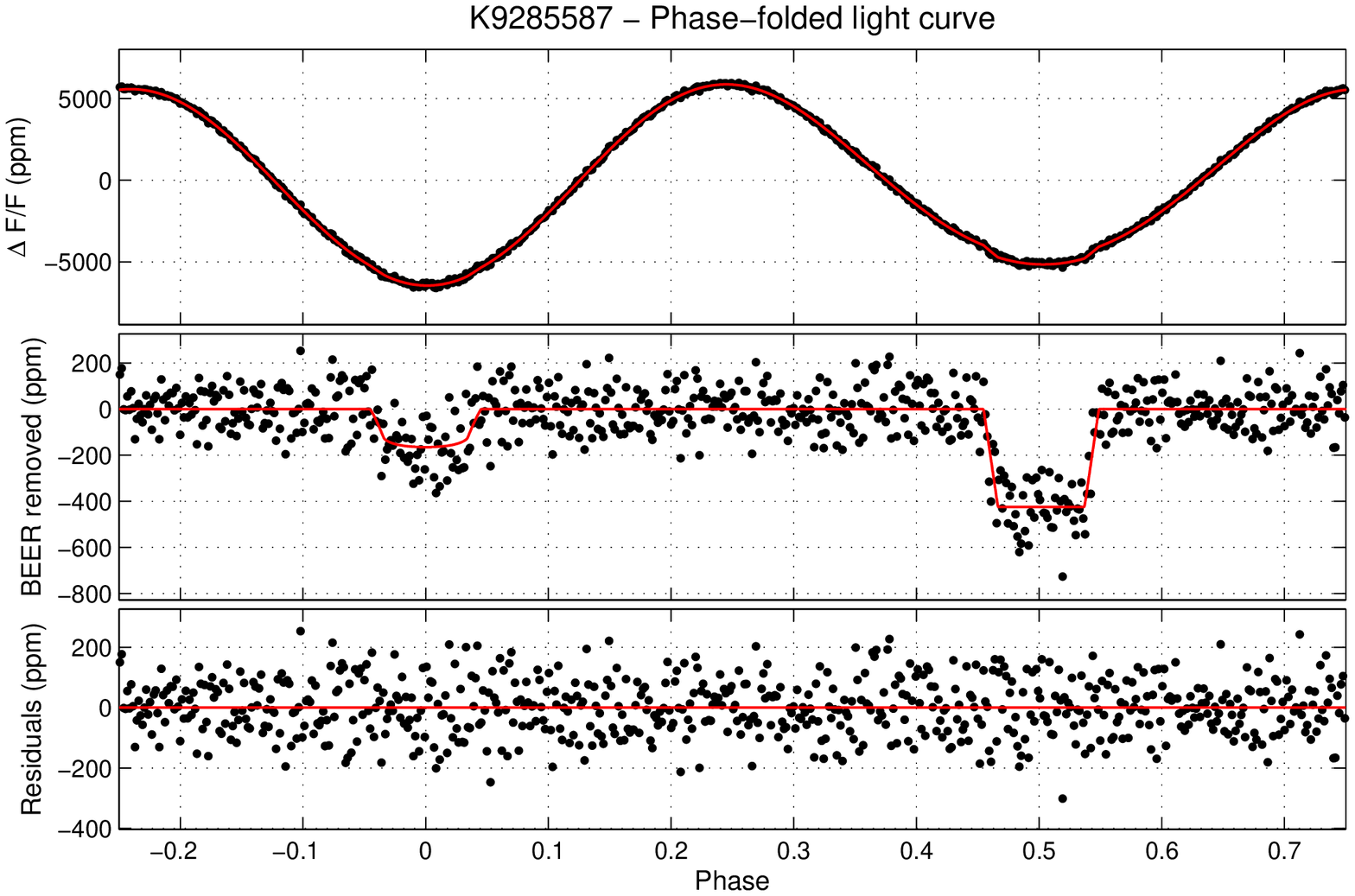}
}
{Fig.~\ref{fig:fold}.--- Continued.}
\label{fig:foldb}
\end{figure*}


\section{Spectroscopic observations}   %
\label{sec:spec}
Follow-up spectroscopic observations of the four stars were obtained with the Tillinghast Reflector Echelle Spectrograph \citep[TRES;][]{furesz08} mounted on the 1.5-m Tillinghast Reflector at the Fred Lawrence Whipple Observatory operated by the Smithsonian Astrophysical Observatory (SAO) on Mount Hopkins in Southern Arizona.
The TRES instrument spans the wavelength range $3850$--$9096$\AA \,, with a nominal resolving power of R$\sim$44,000.
The instrumental setup and observing procedures were as described in \cite{faigler13}. The spectra were extracted and rectified to intensity vs.\ wavelength using the standard procedures developed by \cite{buchhave10}.

For each target, the pseudo-continuum of each spectrum was derived using a large-window $p$-percentile filter \citep{hodgson85}. For such a filter, a value of $p=0.5$ corresponds to a median filtering, while for $p>0.5$ the filter selects flux larger than the median. We empirically chose $p=0.8$ and a filtering window $w$ from the orders' edge to the center of around $200$--$1000$ pixels, in order to obtain the best compromise on continuum matching for both narrow and wide lines. 

The same continuum derivation procedure was also applied to a library of synthetic spectra, with the parameters adjusted visually to $p=0.95$ and $w=500$ pixels, so that the model spectra better fitted the Balmer lines. We later verified that varying these parameters by 10\% had a negligible effect on the derived radial velocities and orbital solutions. The synthetic spectra were also broadened to match the TRES line spread function, on the $Ln(\lambda)$ scale, and assuming a constant R$\sim$44,000 through all orders.

As our targets are hot stars ($T_{\rm eff}>6000$ K) with wide and prominent Balmer lines and narrower Ca II H and K lines and Mg II at 4481 \AA, we tried $\chi^2$ fitting the different lines to the library of synthetic spectra, using an approach similar to \cite{bloemen12}. We did not observe a strong dependence of the surface gravity and metallicity on the $\chi^2$, so we fixed them to fiducial estimates of $\log(g)=4.0$ dex and [Fe/H]=$0.0$. Optimizing the stellar rotation $v \sin i$ over the different lines gave consistent results that are listed in the upper section of Table~\ref{table:RV_results}. We could not constrain the effective temperature using this method, as different lines gave different optimized temperatures with deviations on the order of $1000$--$2000$ K between the lines. We therefore kept 
$T_{\rm eff}$ as a free parameter when finding the optimized synthetic spectrum. 
For our subsequent analysis, described in the next section, we adopted the most recent KIC effective temperatures of Q1--Q17 DR 24 from the \citet{kepexoplanets}, which were determined using a variety of methods \citep{huber14}. 

Radial velocities were derived by cross-correlating the multi-order spectra of each target with the synthetic spectrum that gave the highest peak correlation value \citep[e.g.,][]{mazeh92,zucker03,talor15} chosen from the PHOENIX library of synthetic spectra \citep{hauschildt99}. 
 We excluded all orders with telluric lines and known problems (e.g., broad line on the order's edge, low signal-to-noise ratio, no lines). The resulting RV points are listed in Table~\ref{tab:RV_data}.

\begin{deluxetable}{lr|lr}
\tabletypesize{\scriptsize}
\tablecaption{Radial Velocities of the Four Systems}
\tablewidth{0pt}
\tablehead{
\colhead{Time (BJD$-2450000$)}  & \colhead{RV (km s$^{-1}$)} &  \colhead{ Time (BJD$-2450000$)} & \colhead{RV (km s$^{-1}$)} 
}
\startdata
\tableline
KIC 4169521:           &                  &  KOI-3818: & \\
\tableline
$7092.999960$&$ 11.7(1.0)$ &$6610.588231$&$ 20.0(1.3)$ \\
$7113.939043$&$ 22.3(1.0)$ &$6621.603852$&$ 14.8(1.4)$ \\
$7123.933711$&$- 26.73(93)$ &$7079.014413$&$ 0.1(1.8)$  \\
$7143.946135$&$- 17.31(87)$ &$7091.986663$&$ 26.4(1.4)$  \\
$7146.896942$&$ 17.63(93)$ & $7092.975399$&$- 1.8(2.0)$\\
$7150.931177$&$- 17.72(94)$ & $7094.986119$&$ 20.6(1.3)$\\
$7152.914603$&$- 3.0(1.0)$ &$7110.958442$&$ 20.6(2.1)$  \\
$7153.931701$&$ 16.86(75)$ &$7114.916440$&$ 24.3(1.1)$ \\
                    &                  & $7116.952376$&$- 7.1(1.4)$  \\
                    &                  & $7117.944316$&$ 17.2(1.2)$ \\                                                           
\tableline
KIC 2851474:           &                  &  KIC 9285587: & \\
\tableline
$7087.024901$&$- 16.87(43)$ &$7140.884898$&$ 2.72(93)$  \\
$7095.000035$&$- 22.33(53)$ &$7143.872021$&$- 29.67(60)$ \\ 
$7143.843575$&$ 6.54(29)$ &$7149.956001$&$- 0.35(83)$  \\
$7145.960435$&$ 9.28(35)$ &$7150.959942$&$- 37.33(99)$  \\
$7149.829697$&$- 8.03(47)$ &$7152.946544$&$- 29.59(79)$  \\
$7150.842479$&$- 13.38(53)$ &$7165.918995$&$- 11.50(71)$ \\ 
$7153.903532$&$- 3.81(28)$ &$7170.789666$&$- 38.18(67)$  \\
& & $7175.763105$&$- 20.56(62)$  \\
\enddata
\label{tab:RV_data}
\end{deluxetable}

The radial velocities were used to derive orbital solutions, while taking into account the photometric period and primary eclipse time estimates and their uncertainties.
 This was done by adding the $\chi^2$ of the photometric period and ephemeris to the $\chi^2$ of the RV points.
 To find the orbital solution we searched the parameter space for $\chi^2_{\rm min}$, the minimum combined $\chi^2$. The uncertainties of the parameters were derived by calculating the locus of $\chi^2=\chi^2_{\rm min}+\Delta$, where $\Delta$ corresponds to 1$\sigma$ uncertainty.
 In order to get more realistic uncertainties for the orbital parameters, we inflated only the RV uncertainties by a common multiplicative factor, so that the reduced $\chi^2$ became unity. 

\begin{figure*} [h!]
\centering
\resizebox{17cm}{9cm}
{
\includegraphics{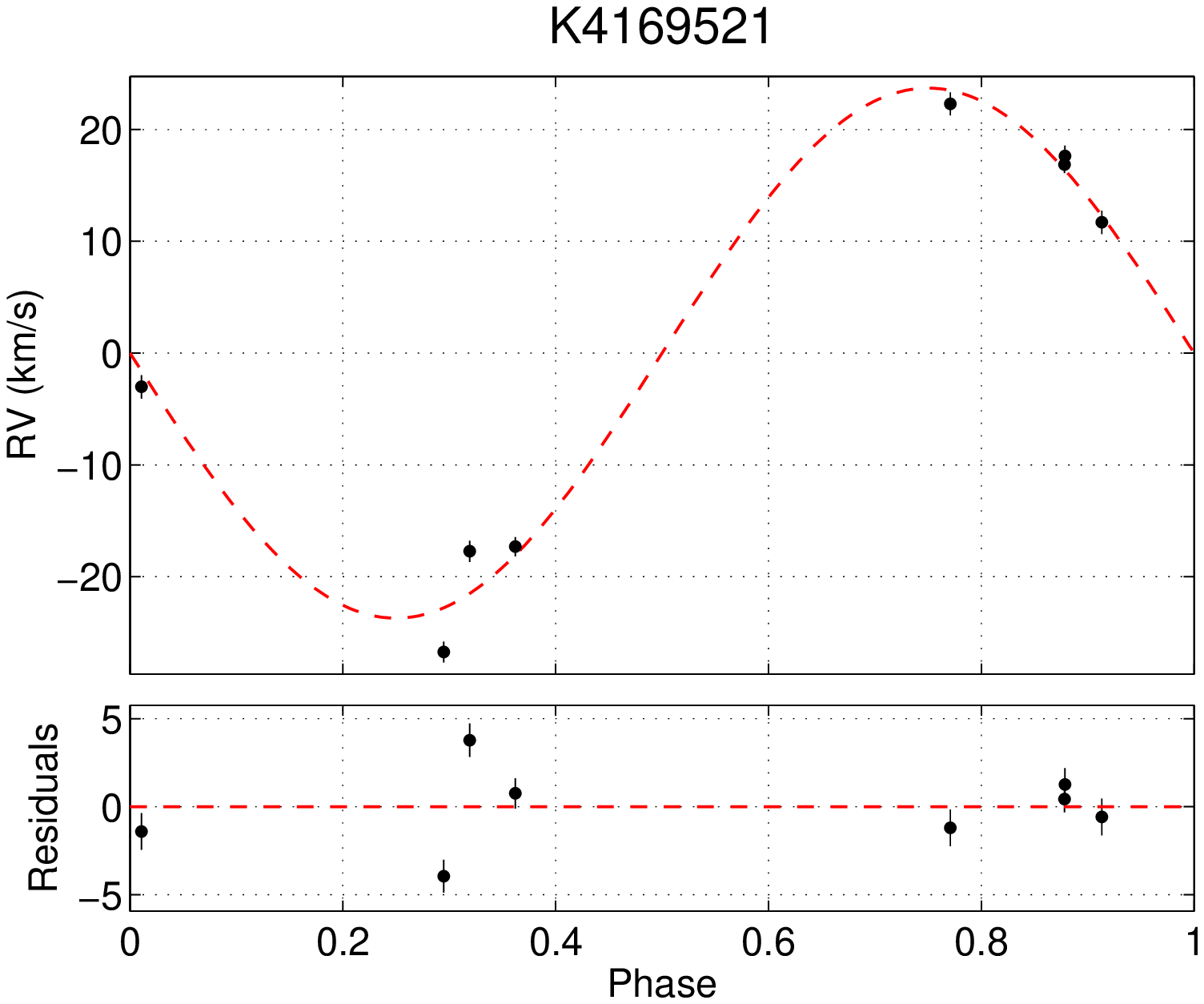}
\includegraphics{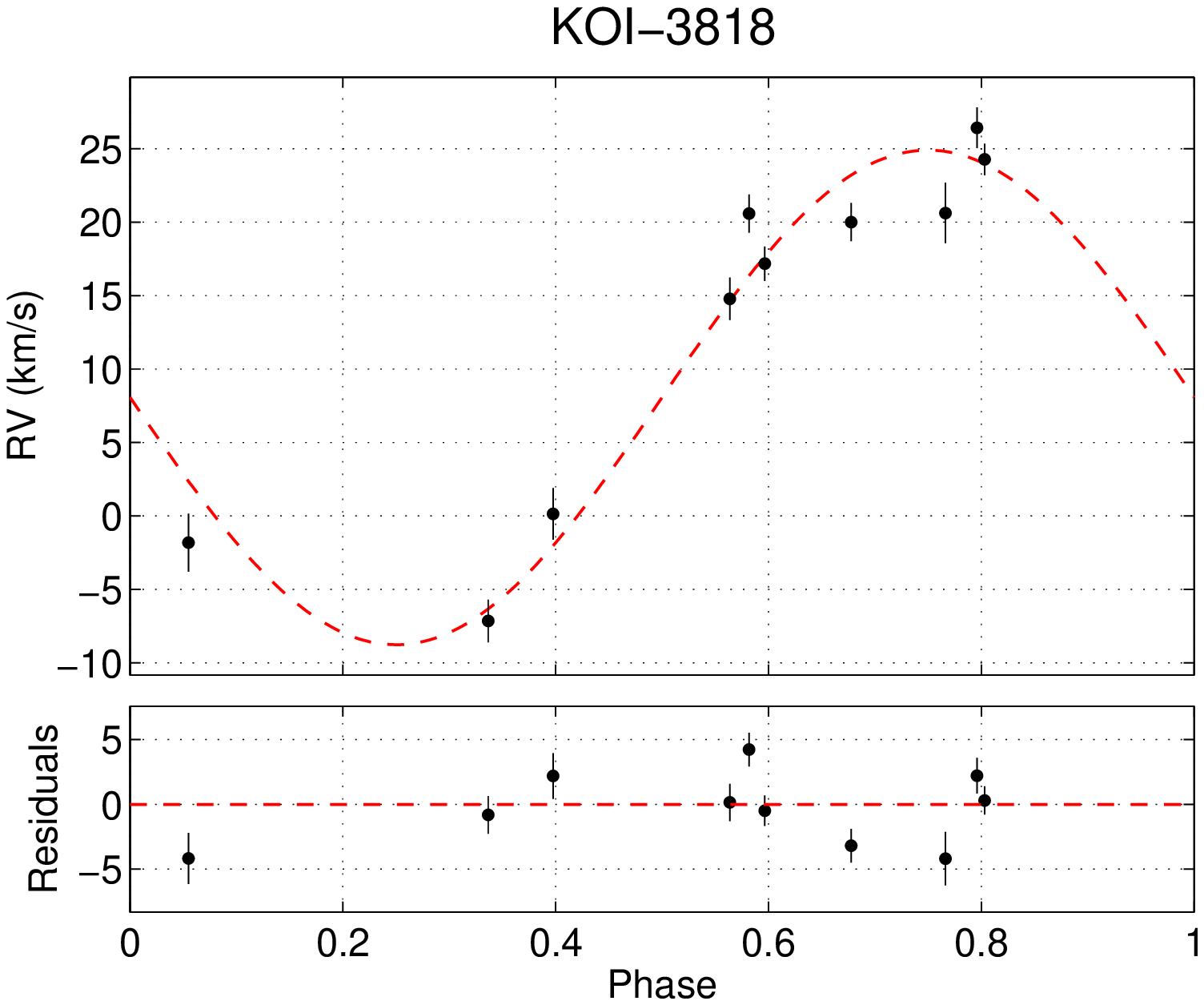}  
}

\centering
\resizebox{17cm}{9cm} 
{
\includegraphics{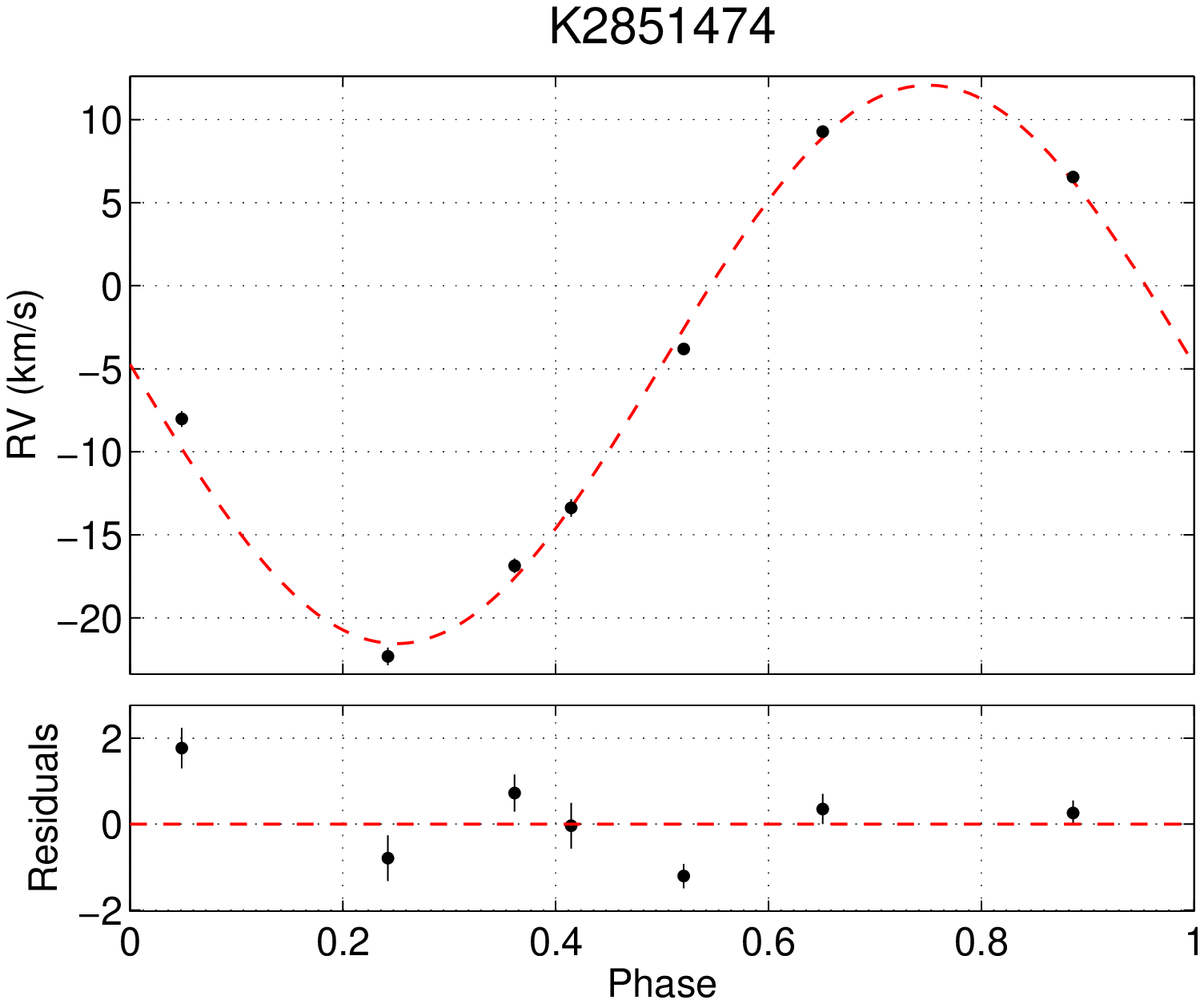}  
\includegraphics{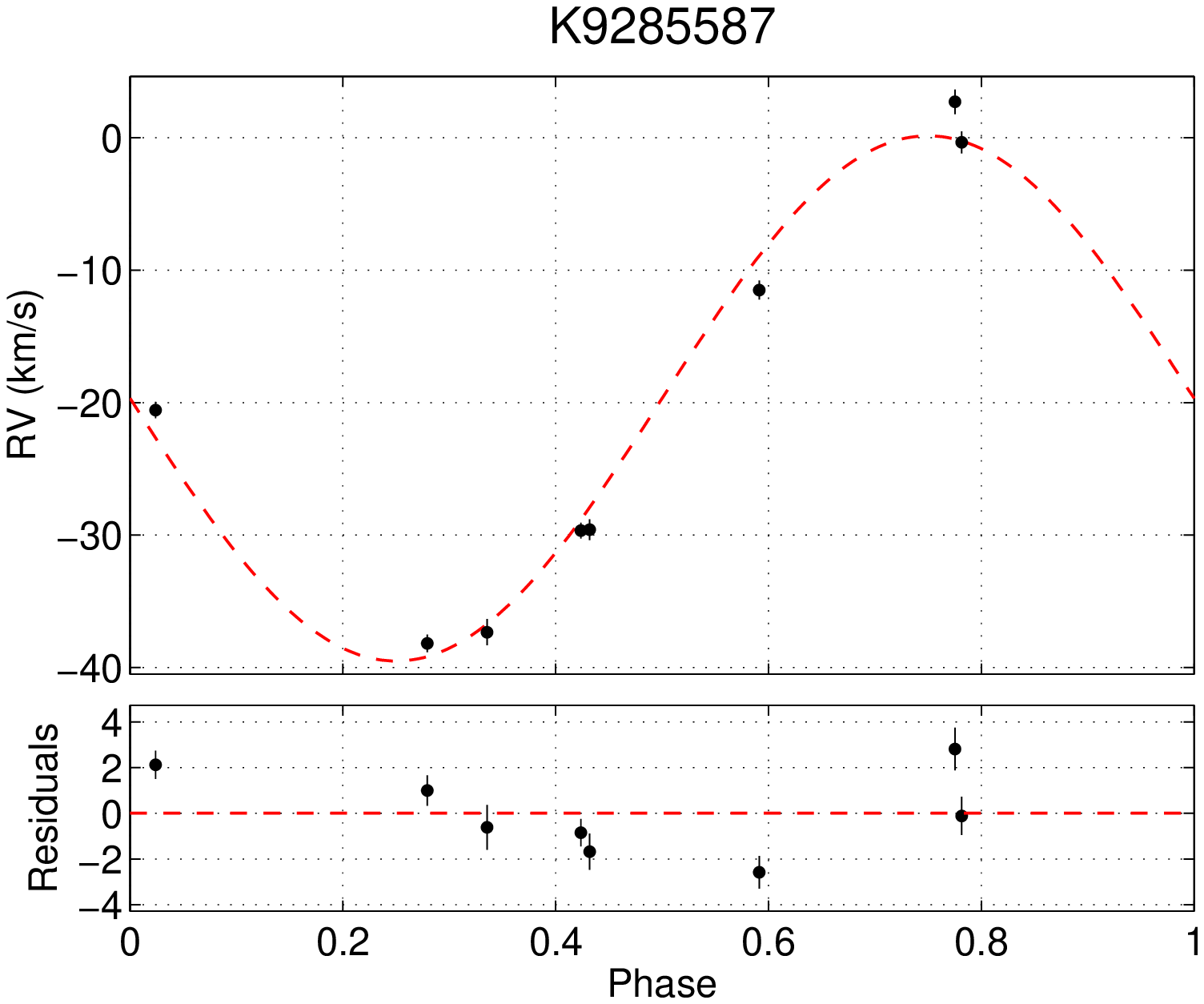}
}
\caption{
RV data and solutions of the four systems.
Dashed lines present the circular velocity curve solutions using the orbital elements listed in Table~\ref{table:RV_results}. Note the different scales of the lower panels of residuals. 
}
\label{fig:RV}
\end{figure*}

We first derived eccentric orbital solutions, but these were not found by an F-test to be preferable over circular-orbit solutions, so here we present the circular-orbits solutions.
The orbital elements of the four systems are listed in Table~\ref{table:RV_results}, and the RV points and orbital models are presented in Figure~\ref{fig:RV}. 

\newcommand{\KOIvsini} {130}

\begin{deluxetable}{lrrrrl}
\tabletypesize{\scriptsize}
\tablecaption{Stellar Properties and RV Orbital Elements}
\tablewidth{0pt}
\tablehead{
& \colhead{KIC 4169521}  & \colhead{KOI-3818} &  \colhead{KIC 2851474} & \colhead{KIC 9285587}
}
\startdata
R.A. & 19h 37m 32.02s  & 19h 18m 57.40s & 19h 24m 59.78s & 19h 34m 59.84s & J2000 R.A. \\
Decl & 39d 15m 18.94s & 41d 57m 54.97s & 38d 02m 33.65s & 45d 45m 42.26s& J2000 Decl \\
$K_{\rm p}$ (mag)& $13.4$ & $11.7$ & $12.6$ & $12.9$ & {\it Kepler} magnitude\\
$v \sin i$ ($\rm km\ s^{-1}$)& $60(15)$ & $\KOIvsini (15)$ & $32(10)$  & $65(15)$ & Projected rotation velocity\\
\tableline
$P$ (d)      &  $1.172555673(89)$  & $3.8170427(48)$ & $2.7682949(87)$ & $1.8119598(58)$ & Orbital period\\
$T$ (BJD-2457000) & $113.91450(15)$   & $91.8111(19)$ & $146.2347(50)$ & $140.8392(58)$ & Maximum RV time \\
$K$ ($\rm km\ s^{-1}$)    & $23.7(1.5)$  & $16.8(1.7)$ & $16.81(90)$ & $19.8(1.4)$ & RV semi-amplitude\\
$V_\gamma$ ($\rm km\ s^{-1}$) & $0.0(1.1)$  & $8.1(1.3)$  & $- 4.74(55)$ & $- 19.68(95)$ & Systemic velocity\\
$f(m_2)$  ($M_\odot$)&  $0.00162(30)$  &  $0.00189(58)$ & $0.00136(22)$ & $0.00146(31)$ & Mass function\\
\tableline
$N_{\rm RV}$             &  $8$  & $10$ & $7$ & $8$ & Number of RV points \\
$F_{\rm RV\_err}$              & $2.59$  & $1.93$ & $2.84$ & $2.77$ & RV uncertainties rescale factor \\
\tableline
\enddata
\label{table:RV_results}
\end{deluxetable}

\section{Modeling the light curves}
\label{sec:model}
For a more complete photometric analysis of these eclipsing WD companion systems we re-analyzed the {\it Kepler} data while trying to minimize the data preparation stages. We first removed data segments with instrumental artifacts \citep{faigler13}, and subtracted a third light constant from each quarter's data using its light curve crowding ratio \citep{jenkins10}. Removal of the long-term trend, periodic modulations, and outliers was performed for each {\it Kepler} quarter by a {\it single} simultaneous robust linear fit \citep{holland77} to the data after masking out the eclipses. The simultaneous fit was to four sets of functions: long-term cosine-detrend functions of periods down to a minimum of twice the orbital period \citep{mazeh10}, BEER cosine and sine functions of the first four orbital-period harmonics, jump functions at predefined {\it Kepler} times \citep{faigler13}, and high-frequency stellar-activity sine and cosine functions.

The high-frequency functions were incorporated only for KOI-3818 and KIC 9285587, for which the light curve and amplitude spectrum show significant high-frequency modulations.
The high-frequencies list of KOI-3818 was constructed manually because its modulation spectrum showed a simple structure of peaks at $\sim$$2.113$ cycles day$^{-1}$ and its harmonics. Consequently, we included the $\sim$$2.113$ cycles day$^{-1}$ frequency  in the list, along with its three first harmonics, and added two and one side-lobe frequencies on each side of the base frequency and its first harmonic, respectively. The side-lobe frequencies were separated by $1/90$ cycles day$^{-1}$, which is the natural frequency separation for a fit performed on a single {\it Kepler} quarter of a typical duration of $90$ days.

The high-frequency modulation spectrum of KIC 9285587 showed no clear structure, so we constructed its frequency list through a pre-whitening iterative process. At each iteration, we fitted the data to sine and cosine functions of the previous frequency list, and derived the spectrum of the fit residuals. The frequency of the highest residuals-spectrum peak was then added to the previous list, and the updated list was used in the next iteration. This process was stopped based on a Bayesian information criteria.

 The resulting frequency lists of KOI-3818 and KIC 9285587 were then used in the simultaneous robust fit of each {\it Kepler} quarter data to the detrend, jumps, BEER, and high-frequency functions. As the fit results we report an amplitude and its uncertainty as that amplitude's median and its median absolute deviation across the {\it Kepler} quarters, respectively \citep{faigler15}.
 The fitted high frequencies and amplitudes of these two systems are listed in Table~\ref{tab:freqs}. For each frequency, the listed amplitude is the 2-norm of that frequency's sine and cosine fitted amplitudes (i.e. $\sqrt{A_{\rm cos}^2+A_{\rm sin}^2}$, where $A_{\rm cos}$ is the cosine amplitude, and $A_{\rm sin}$ is the sine amplitude, of that frequency).
 It may seem strange that for most frequencies the fitted amplitudes are much smaller than their uncertainties. However, this only means that the amplitudes vary wildly across the quarters, which is indicative of the non-periodic nature of these modulations. 

\begin{deluxetable}{lr|lr}
\tabletypesize{\tiny}
\tablecaption{Fitted Frequencies and Amplitudes of High-Frequency Modulations}
\tablewidth{0pt}
\tablehead{
\colhead{Frequency (cycles day$^{-1}$)}  & \colhead{Amplitude (ppm)} &  \colhead{Frequency (cycles day$^{-1}$)} & \colhead{Amplitude (ppm)} 
}
\startdata
\tableline
KOI-3818:           &                  &  KIC 9285587: & \\
\tableline
$2.090690$ & $7 \pm 35$ & $18.948908$ & $99 \pm 454$ \\
$2.101801$ & $11 \pm 70$ &$19.091935$ & $198 \pm 633$ \\
$2.112913$ & $93 \pm 560$ &$20.195763$ & $249 \pm 1685$ \\
$2.124024$ & $16 \pm 106$ &$20.693235$ & $151 \pm 292$ \\
$2.135135$ & $12 \pm 44$ &$20.941710$ & $44 \pm 438$\\
$4.214714$ & $10 \pm 18$ &$21.006874$ & $22 \pm 635$ \\
$4.225825$ & $101 \pm 322$ & $21.031024$ & $689 \pm 1486$ \\
$4.236936$ & $9 \pm 29$ &$21.596262$ & $51 \pm 162$ \\
$6.338738$ & $15 \pm 41$ &$22.045434$ & $189 \pm 893$ \\
$8.451650$ & $4 \pm 31$ &$22.134748$ & $192 \pm 653$ \\
& & $22.880592$ & $187 \pm 493$ \\
& & $23.362137$ & $105 \pm 106$ \\
& & $23.493714$ & $96 \pm 284$ \\
& & $23.611342$ & $95 \pm 732$ \\
& & $24.052498$ & $205 \pm 1236$ \\
& & $24.341467$ & $35 \pm 222$ \\
                                      
\enddata
\label{tab:freqs}
\end{deluxetable}
\clearpage

  For the shortest period system, KIC 4169521, we had to fit {\it five} BEER harmonics in order to get a good fit to the data. This is probably due to the significance of higher orders of the ellipsoidal modulation, resulting from the proximity of the two binary objects. The simultaneous robust fit was performed after taking the logarithm of the data to account for the multiplicative nature of the eclipses and the phase modulations \citep{huang13}. 
The fitted BEER amplitudes, with their uncertainties derived from the quarter-to-quarter scatter \citep{faigler15}, are listed in Table~\ref{table:sys}.

Next, we subtracted the fitted trend model from the {\it unmasked} data and analyzed the detrended eclipses.
For that we ran a Markov chain Monte Carlo (MCMC) analysis, using the Ensemble Samplers method \citep{goodman10} that is invariant to affine transformations of the parameter space, making it much more efficient for problems with correlated parameters.
The MCMC analysis was performed by simultaneously fitting the primary and secondary eclipses using a {\it Kepler} long-cadence integrated \citet{mandel02} model with quadratic limb darkening, assuming a circular orbit. The model limb-darkening coefficients could not be constrained from the data, so we estimated their values and uncertainties by interpolating the \citet{claret11} limb-darkening tables, using the KIC effective temperatures and the fiducial estimates of $\log(g)=4.0$ dex and [Fe/H]=$0.0$.

The model we fitted to the data had 8 free parameters:  orbital period $P$, middle of primary eclipse time $T_0$, radius ratio $R_2/R_1$, scaled separation $a/R_1$, flux ratio $F_2/F_1$, impact parameter $b$, secondary-eclipse time shift $\Delta T_{\rm sec}$, and secondary-to-primary eclipses durations ratio $\tau_{\rm sec}/\tau_{\rm prim}$. The secondary-eclipse time shift was added to the model in order to account for the R\o mer delay \citep{kaplan10,bloemen12} and the $e\cos \omega$ component of a possible small eccentricity, while the secondary-to-primary eclipses durations ratio was added to account for the $e\sin \omega$ component of such a possible eccentricity.

Figure~\ref{fig:fold} presents the cleaned and detrended binned data and the best-fit \citet{mandel02} model combined with, and without, the BEER phase modulation, folded at the orbital period. 
The three upper sections of Table~\ref{table:sys} summarize the results of the photometric analysis. The first section of the table lists the third light fraction and the 
limb-darkening coefficients; the second lists the MCMC medians and $1 \sigma$ uncertainties of the model parameters; and the third section lists the BEER amplitudes with their uncertainties.

%
\begin{deluxetable}{lrrrrl}
\tabletypesize{\scriptsize}
\tablecaption{System Parameters}
\tablewidth{0pt}
\tablehead{
& \colhead{KIC 4169521}  & \colhead{KOI-3818} &  \colhead{KIC 2851474} &\colhead{KIC 9285587}
}
\startdata
$f_{3}$  & $0.042 $ & $0.017$ & $0.109$ & $0.027$ & Average {\it Kepler} third light fraction \\
$\gamma_{1}$  & $0.274(19) $ & $0.230(16)$ & $0.254(20)$ & $0.258(27) $ & linear limb-darken coeff. \\
$\gamma_{2}$  & $0.317(21)$ & $0.2919(74) $ & $0.297(17)$ & $0.305(28)$ & quad. limb-darken coeff. \\
\tableline
P (days)& $1.172555671(69)  $ & $3.8170428(39)$ & $2.7682925(66)$ & $1.8119579(48)$ & Orbital period \\
$T_0$ ($\rm BJD$-$2454833$) & $863.598047(83)$ & $862.73788(73) $ & $907.6432(17)$ & $864.1697(30)$ & Middle of primary eclipse time \\
$R_2/R_1$ & $0.04131(22)$ & $0.01256(22) $ & $0.01118(31)$ & $0.01206(57)$ & Radius ratio \\
$a/R_1$ & $2.704(26) $ & $6.85(75)$ & $3.69(56)$ & $3.77(62)$ & Normalized separation \\
$F_2/F_1$& $0.0053785(52)$ & $0.0002546(24) $ & $0.0002227(35)$ & $0.000425(11)$ & {\it Kepler}-band flux ratio \\
$b$ & $0.8564(19) $ & $0.31(0.22)$ & $0.37(0.26)$ & $0.38(0.26)$ & Impact parameter \\
$\Delta T_{\rm sec}$ (minutes)& $0.02(12)$ & $6.0(1.2)$ & $0.7(2.7) $ & $4.6(4.5)$ & Secondary eclipse time delay  \\
$\tau_{\rm sec}/\tau_{\rm prim}$ & $0.9921(39)$  & $1.001(10)$ & $0.976(18)$ & $1.039(48)$ & Eclipses durations ratio\\

\tableline
\multicolumn{6}{l}{BEER harmonics semi-amplitudes (ppm):} \\
$a_{1c}$  & $818.7(8.9)  $ & $13.0(4.5) $ & $- 39(23) $ & $- 312(13) $ & $\cos \phi$ semi-amplitude \\
$a_{1s}$ (beaming) & $78.5(7.3)  $ & $132.3(3.4) $ & $126(11) $ & $138(16) $ & $\sin \phi$ semi-amplitude \\
$a_{2c}$ (ellipsoidal) & $- 8402(18) $ & $- 621.8(3.7) $ & $- 2296(157) $ & $- 5601(30) $ & $\cos 2\phi$ semi-amplitude \\
$a_{2s}$ & $67.0(6.3) $ & $45.9(3.3) $ & $- 112(16) $ & $- 153(78) $ & $\sin 2\phi$ semi-amplitude \\
$a_{3c}$ & $- 738.1(6.0) $ & $- 26.4(1.7)  $ & $- 139(14) $ & $- 465(13) $ & $\cos 3\phi$ semi-amplitude \\
$a_{3s}$ & $- 4.7(4.7)  $ & $4.7(2.8)  $ & $- 4.5(3.1) $ & $- 23(12)  $ & $\sin 3\phi$ semi-amplitude \\
$a_{4c}$ & $207.0(7.8)  $ & $10.3(2.1)  $ & $37.8(3.0) $ & $94.9(9.0)  $ & $\cos 4\phi$ semi-amplitude \\
$a_{4s}$ & $- 5.2(3.9)   $ & $- 0.3(2.6) $ & $0.9(3.3) $ & $7.1(6.2)  $ & $\sin 4\phi$ semi-amplitude \\

\tableline
$\alpha_{\rm beam}$  & $0.643(49)$ & $0.570(34) $ & $0.614(47) $ & $0.650(52) $ & Beaming factor\\
$K_{\rm beam}$ ($\rm km\ s^{-1}$)  & $9.1(1.1)$ & $17.4(1.2)$ & $15.4(1.9)$ & $15.9(2.4)$ &RV semi-amplitude from beaming\\
$q_{\rm ellip}$  & $0.1191(51)$ & $0.144(42)$ & $0.078(28)$ & $0.191(76)$ & Mass ratio from ellipsoidal\\

\tableline
$T_{\rm eff}$ (K)  & $8290^{+250}_{-320}$& $9170^{+280}_{-390}$ & $8580^{+240}_{-370}$ & $8230^{+230}_{-350}$ & KIC effective temperature\\
$M_1$ $(M_\odot)$& $1.982(92)$ & $2.14(12)$ & $2.34(19) $ & $1.94(16)$ &  Primary mass\\
$R_1$ $(R_\odot)$ & $2.247(40)$ & $1.99(28) $ & $3.06(64)$ & $2.13(48)$ &  Primary radius\\
$\rho_1$ $({\rm g/cc})$ & $0.2460(72)$ & $0.38(11)$ & $0.114(44)$ & $0.28(12)$ &  Primary density\\
Age (Gyr) & $0.74(0.13)$ & $0.42(0.1)$& $0.6(0.1)$ & $0.72(0.14)$ & Age \\
$K_{\rm RV}$ $(\rm km\ s^{-1})$ & $23.7(1.5)$  & $16.8(1.7)$ & $16.81(90)$ & $19.8(1.4)$ & Spectroscopic RV semi-amplitude\\
$P_{\rm rot}$ (days) & $1.79(60) $  & $0.79(14)$ & $5.0(2.4)$ & $1.71(62)$ & Primary rotation period\\

\tableline
$T_2$ (K)& $12114(629)$ & $10854(617)$ & $10306(545)$ & $11748(707)$ & Secondary temperature  \\
$M_2$ $(M_\odot)$& $0.210(15)$ & $0.220(26)$ & $0.210(18)$ & $0.191(19)$ &  Secondary mass\\
$R_2$ $(R_\odot)$ & $0.0929(19)$ & $0.0260(39)$ & $0.0346(83)$ & $0.0260(68)$ &  Secondary radius\\
$\rho_2$ $({\rm g/cc})$ & $372(28)$ & $19388(6470)$ & $7365(3265)$ & $15847(7559)$ &  Secondary density\\
$q $ & $0.1064(74) $ & $0.103(11)$ & $0.0894(54)$ & $0.0979(76)$ &  Mass ratio\\

\tableline
$\Delta T_{\rm R\o mer}$ (minutes)& $0.3570(77)$ & $0.856(24)$ & $0.726(26)$ & $0.506(12)$ &R\o mer delay \\
$e \cos \omega $ & $0.00030(11)$ & $- 0.00146(33) $ & $0.0000(11)$ & $- 0.0025(27)$ &  Eccentricity cosine component\\
$e \sin \omega $ & $0.0039(20)$ & $- 0.0004(50) $ & $0.0123(90)$ & $- 0.019(23)$ &  Eccentricity sine component\\

\enddata
\label{table:sys}
\end{deluxetable}

\section{The parameters of the systems}
\label{sec:params}

\subsection{Masses and radii}
\begin{figure*} [h]
\centering
\resizebox{14cm}{12cm} 
{
\includegraphics{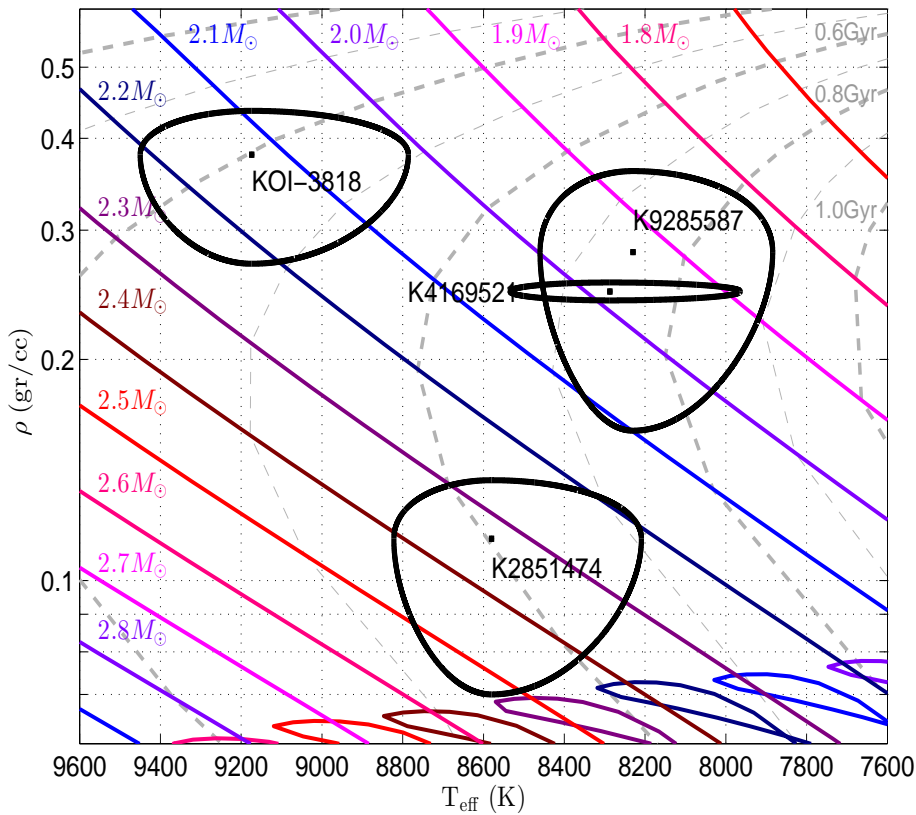}
}
\caption{
Location of the four systems on Dartmouth density-temperature plane stellar evolutionary tracks. Black contours show the $1$$\sigma$ uncertainties' boundaries.
}
\label{fig:evo}
\end{figure*}

To estimate the primary mass, we used a method similar to the one used by \cite{rappaport15}. We first estimated the primary density by using Kepler's third law with the photometric period, the scaled separation $a/R_1$, and a rough initial estimate of $0.1$ for the mass ratio \citep{seager03}. We then used the Dartmouth stellar evolution tracks \citep{dotter08} in the mean stellar density $\rho$ and effective temperature $T_{\rm eff}$ plane, to estimate the primary mass. Next, using the RV mass function, we derived the secondary mass and thus got a better estimate for the mass ratio. We repeated this process with the updated mass ratio until the primary and secondary mass estimates converged to stable values, which required only two iterations for each of the four systems. To test this process, we also tried initial mass-ratio values of $0.2$ or $0.05$, and in both cases it converged to the same final parameters estimates within only two iterations. Figure~\ref{fig:evo} presents the location of the four systems on stellar evolution tracks, for stars of a range of initial masses, in the density-temperature plane.

Using the masses of the two components of the binary we then derived the semimajor axis and the two radii using Kepler's third law and the photometry derived parameters. The WD radius $R_2$ was estimated while also taking into account the gravitational-lensing effect on the primary eclipse depth \citep{marsh01,agol03,bloemen11,muirhead13,kruse14}. This effect is expected to be almost negligible for short-period low-mass WD companions. Indeed, the effect was most prominent in our longest period system, KOI-3818, for which it inflated the estimated WD radius by only $\sim 4\%$, or $\sim 0.4 \sigma$.   

\subsection{Photometric beaming RV}
\label{sec:kbeam}
For a negligible luminosity companion, the beaming phase modulation amplitude is proportional to the primary RV amplitude, with a proportionality constant of
$4 \alpha_{\rm beam} / c$, 
where $c$ is the speed of light, and $\alpha_{\rm beam}$ is mainly a function of the primary effective temperature \citep{loeb03,zucker07,bloemen11}. We note that 
$\alpha_{\rm beam}=\frac{3-\alpha}{4}=\frac{\langle B \rangle}{4}$, 
where $\alpha$ is the power-law index used by \citet{loeb03} and $\langle B \rangle$ is the photon weighted bandpass-integrated beaming factor used by \citet{bloemen11}. Using this relation, we estimated the expected RV semi-amplitude from the photometric beaming semi-amplitude, $K_{\rm beam}$, following the same method used by \cite{faigler13}.

For KOI-3818, KIC 2851474, and KIC 9285587, our $K_{\rm beam}$ estimates are consistent with the spectroscopic RV estimates, $K_{\rm RV}$, with deviations of $0.3 \sigma$, $0.7 \sigma$ and $1.4 \sigma$ respectively. 
For KIC 4169521 our $K_{\rm beam}$ estimate is significantly lower than the spectroscopic $K_{\rm RV}$. This may be explained by the fact that for this system the orbital-period cosine reflection-phase-modulation is more than $10$ times larger than the corresponding sine modulation. 
In such a case, a small phase shift in the cosine reflection modulation, due to stellar atmospheric effects, can significantly modify the sine phase modulation semi-amplitude, which we interpret as the beaming semi-amplitude. Similar effects were previously observed for the WD secondary system KIC 9164561 \citep{rappaport15} and for the hot-Jupiter system Kepler-76 \citep{faigler13}. 

\subsection{Photometric ellipsoidal mass ratio and amplitudes}
In theory, for a co-rotating primary the mass ratio can be photometrically derived from the ellipsoidal amplitude combined with the scaled separation $a/R_1$, the inclination, and the estimated effective temperature \citep[Eq.~1 in][]{morris93,zucker07}. However, in practice this calculation usually results in large uncertainties that give little meaning for the resulting mass ratio. In addition, it has been shown theoretically and through observations in multiple cases that for a massive star with a radiative envelope, or for asynchronous primary rotation, the ellipsoidal derived mass ratio can be significantly different from the real one \citep[e.g.,][and references therein]{pfahl08,carter11,bloemen12}. For these reasons we chose not to use the ellipsoidal modulation amplitude in order to derive the systems' parameters. It is interesting, though, to compare the ellipsoidal derived mass ratio $q_{\rm ellip}$, with the mass ratio we derive from the combination of primary and secondary eclipses fitting, RV solutions, and stellar evolution models. For KOI-3818, KIC 2851474, and KIC 9285587, the ellipsoidal derived mass ratios are consistent with our mass-ratio estimates, at the level of $1.0$$\sigma$, $0.7$$\sigma$, and $1.1$$\sigma$ respectively. We note, though, that this is merely due to the large relative uncertainty of $29\%$--$40\%$ in the derived $q_{\rm ellip}$ of these systems, making it of little importance. For KIC 4169521, $q_{\rm ellip}$, with a much smaller relative uncertainty of $\sim$$4\%$, is still consistent with our derived mass ratio at the $1.2$$\sigma$ level. 
This result, which assumes a co-rotating primary, fits well with the 
consistency of the orbital period and the stellar rotation period of KIC 4169521 (Section \ref{sec:rot}).

It is also interesting to derive the expected first four BEER amplitudes using Eq.~1 of \cite{morris93}, and compare them to the amplitudes measured from the data. Again, for KOI-3818, KIC 2851474, and KIC 9285587, the resulting uncertainties are too high to provide any meaningful result. For KIC 4169521, however, for the 2nd, 3rd and 4th cosine amplitudes ($a_{2c}$, $a_{3c}$ and $a_{4c}$) we find
 $-8070 \pm 820$, $-750 \pm 90$, and $213 \pm 27$ ppm, respectively. These are very consistent with the BEER amplitudes derived from the light curve (Table~\ref{table:sys}), which again supports the co-rotation scenario of this system.
 
\subsection{Stellar rotation period}
\label{sec:rot}
In general, given enough time, tidal interaction in short-period binaries leads to synchronization, circularization and alignment of the system \citep[e.g.,][]{mazeh08}. The data that we have at hand allow us to check if our four dA+WD reached such a stable dynamical configuration. If true, we can assume alignment and test synchronization by using the spectroscopic projected rotation velocity $v\sin i$, with the derived impact parameter and stellar radius, to estimate the stellar rotation period. 
Table~\ref{table:sys} lists the resulting stellar rotation periods of the four systems, under these assumptions.
For KIC 4169521, KIC 2851474, and KIC 9285587, the resulting rotation periods are consistent with the orbital periods, but this may be attributed to their large relative uncertainty of $\sim$$40\%$. For KOI-3818 this exercise gives a rotation period of $0.79 \pm 0.14$ days, far from being consistent with the $3.82$ days orbital period, suggesting a fast-rotating primary. 

\subsection{Eccentricities}
\label{sec:ecc}
In addition, the precise {\it Kepler} data allow us to look for small orbital eccentricity, another aspect of the tidal interaction. This is done by deriving the secondary-eclipse time shift $\Delta T_{\rm sec}$, and the secondary-to-primary eclipses durations ratio $\tau_{\rm sec}/\tau_{\rm prim}$. We estimated $e \cos \omega$ from the secondary-eclipse time shift by taking the first order in $e$ of Equation~1 in \cite{dong13}, after subtracting from it the expected R\o mer delay \citep{kaplan10,bloemen12}. Next, $e \sin \omega$ was derived from the secondary-to-primary eclipses durations ratio, using the first order approximation in $e$ of Equation~7 in \cite{tingley05}. 
Again, for KOI-3818 we identify a small but significant eccentricity component of $e \cos \omega=-0.00146\pm33$.
The data of KIC 4169521 are precise enough to rule out even such a small eccentricity, with a $2$$\sigma$ upper limit of $0.0005$.
Unfortunately, the light curves of the other two systems, KIC 2851474 and KIC 9285587, yielded larger upper limits, $0.002$ and $0.008$, respectively, and therefore do not allow us to rule out eccentricities of the order of $0.001$.

\section{Review of the individual systems}
\label{sec:indiv}
In this section we review specific features of each of the four systems.

\subsection{KIC 4169521}
This system is listed in the {\it Kepler} EB catalog with an orbital period that is consistent with what we find, but with a primary eclipse time $\rm BJD_0$ that is shifted by half a period, because the catalog defines $\rm BJD_0$ as the time of the deeper eclipse.
This system's orbital period of $1.17$ days is the shortest among the dA+WD EBs discovered so far in {\it Kepler}. It is, though, within the period range of $0.67$--$2.2$ days of the previously discovered such systems in WASP. It is also the hottest WD with the largest-radius of the four systems reported here. It is then not a surprise that the light curve of this system shows a significant reflection/emission modulation of $+818.7 \pm 8.9$ ppm. This is likely due to light originating from the WD, which is scattered off of or thermally reprocessed and later emitted from the atmosphere of the A star. It is similar to the reflection/emission modulation observed for the WD secondaries systems KIC 10657664, KOI-1224, and KIC 9164561 \citep{carter11,breton12,rappaport15}. Such a modulation, in turn, can significantly modify the derived beaming amplitude, which is likely the case here (see Section~\ref{sec:kbeam}).

\subsection{KOI-3818}
Our analysis detects small but significant orbital eccentricity (Section \ref{sec:ecc}) and fast rotation of the primary, which highly deviates from synchronization (Section \ref{sec:rot}), indicating an incomplete tidal interaction.

In addition, the light curve of this system shows significant modulations at frequencies of $\sim$$2.1$ cycles day$^{-1}$ and its harmonics (Figures~\ref{fig:lc} and \ref{fig:psd}).
With a secondary-to-primary {\it Kepler}-band flux ratio of $\sim 2.5 \times 10^{-4}$, it is safe to assume that the A-star is the source of this variability. 
This is similar to the photometric modulations, termed ``low-frequencies'', seen in many A-type stars in the {\it Kepler} data
  \citep{balona11,balona14,balona15,guzik15}.  
There is an active ongoing research trying to explain the source of such modulations, with no clear conclusions.
Based of the harmonic structure of the modulation, and the consistency of its frequency with our derived stellar rotation period (Section \ref{sec:rot}), we speculate that it may be associated with the A-star rotation.

\subsection{KIC 2851474}
This system appears in the {\it Kepler} EB catalog, with an orbital period and a $\rm BJD_0$ consistent with our findings,
but the catalog folded light curve mainly shows an ellipsoidal variation, with no visible eclipses.
This binary features the most massive primary A-star of the four systems. For the A-star we derive a mass of $2.3 M_{\odot}$ and a radius of $3.1 R_{\odot}$, suggesting that the primary has already started to evolve away from the main sequence. This is also indicated by its position on the stellar evolution tracks illustrated in Figure~\ref{fig:evo}.

\subsection{KIC 9285587}
The light curve of this system shows significant modulations at the $19$--$24$ cycles day$^{-1}$ frequency range (Figures~\ref{fig:lc} and \ref{fig:psd}).
Similar to KOI-3818, the small secondary-to-primary {\it Kepler}-band flux ratio of $\sim 4 \times 10^{-4}$ suggests that the A-star is the source of this variability. 
These $\delta$ Scuti-like pulsations are similar to high-frequency modulations seen in many A-type stars in the {\it Kepler} data \citep{breger00,balona15,guzik15}. 

\section{Summary and Discussion}
\label{sec:disc}
We report the discovery of four short-period dA+WD EBs in the {\it Kepler} light curves. The 4 systems add to the 6 previously known short-period dA+pre-He-WD EBs in the {\it Kepler} data 
 \citep{rowe10,vankerkwijk10,carter11,bloemen12,breton12,matson15,rappaport15}, 
 and the 18 such WASP systems \citep{maxted11,maxted13,maxted14}.
 All of the 6 {\it Kepler} systems and 3 of the WASP systems have accurate measurements of the primary and secondary mass,  radius, and $T_{\rm eff}$.
 The new systems' orbital-period range is $1.17$--$3.82$ days, 
 well within the $0.67$--$23.9$ days range of the previously known systems.
The masses of the WD secondaries of the new systems range from $0.19$ to $0.22$ $M_{\odot}$,
somewhat overlapping the lower edge of the $0.19$--$0.3$ $M_{\odot}$ mass range of the known systems with derived secondary masses. 

Each dA+WD binary that we currently observe has evolved from a system in which the primordial primary was the progenitor of the current WD.  
 Such systems are believed to have gone through a mostly stable mass transfer from the WD progenitor to the current primary star, which gained a significant part of its current mass through this process \citep{podsiadlowski02,panei07,rappaport09,vankerkwijk10}. The WD evolution could involve several hydrogen shell flashes of the degenerate remnant, before it settles on the He WD cooling track \citep{podsiadlowski02,panei07,althaus13}.

\begin{figure*} [h!]
\centering
\resizebox{12cm}{10cm} 
{
\includegraphics{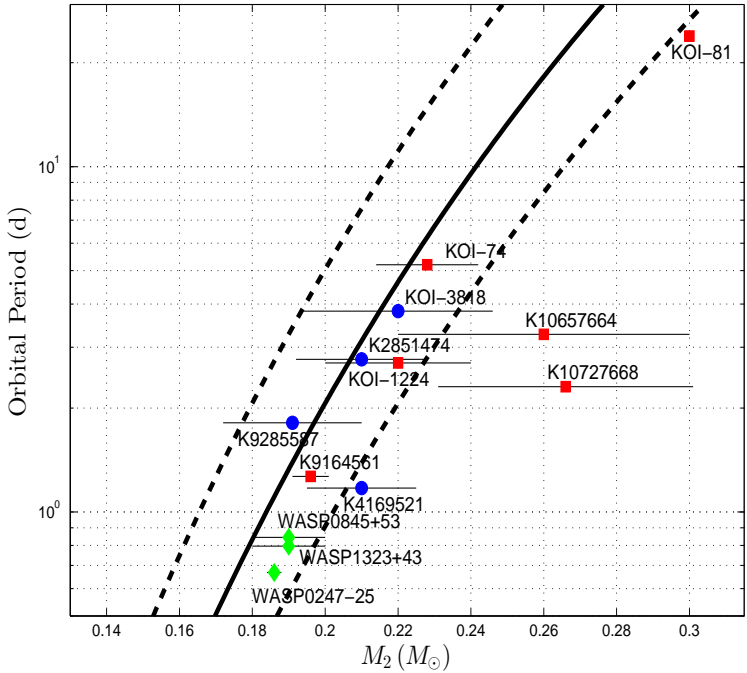}
}
\caption{
WD mass--period relation of the four new and nine previously known systems. The solid curve shows the expected relation from \citet{lin11}, and the dashed lines show the $\pm 10\%$ uncertainties of the model. Red squares are the six previously known {\it Kepler} systems,  green diamonds are the three previously known WASP systems, and blue circles are the four newly discovered systems.
}
\label{fig:PM}
\end{figure*}

Mass-transfer models predict a direct dependence of the final orbital period on the WD mass. To check this dependence, we follow \cite{carter11} and plot in Figure~\ref{fig:PM} the current orbital period as a function of the WD mass of the four new and nine previously known systems with derived masses, on top of the expected mass--period relation from \citet{lin11} \citep[see also e.g.,][]{joss83,savonije83,rappaport95}. Indeed, all systems seem consistent with the mass--period relation, supporting the stable mass-transfer scenario.

It is also interesting to examine the WD radii of the newly discovered systems. For KIC 4169521 we derive a bloated WD radius of $0.09 R_{\odot}$, well within the WD radius range of $0.04$--$0.33$ $R_{\odot}$ of the already known dA+pre-He-WD systems. However, the WD radii of the other three new systems are in the range of $0.026$--$0.035$ $R_\odot$, the smallest WD radii discovered so far in short-period eclipsing dA+WD binaries.
 This is reflected by the very shallow eclipses of these three systems, with depths on the order of $100$--$400$ ppm, 
well below those of the previously detected dA+pre-He-WD systems.

\begin{figure*} [h!]
\centering
\resizebox{14cm}{12cm} 
{
\includegraphics{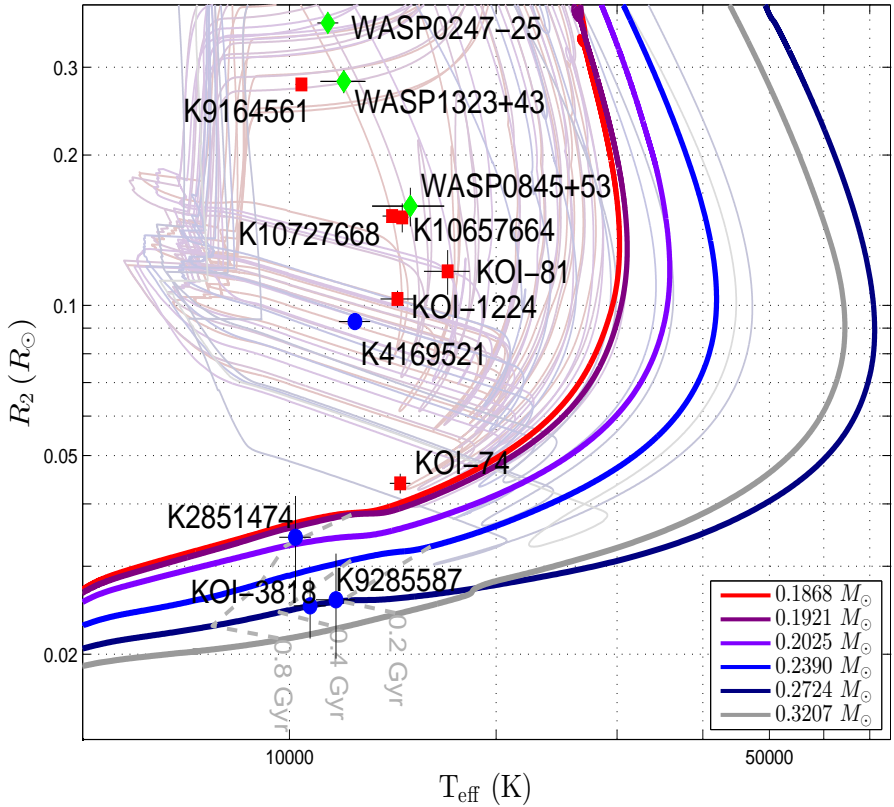} 
}
\caption{
WD evolution tracks for binary systems that have gone through stable mass transfer, for a selected set of WD masses from \citet{althaus13}. Colored curves show the final cooling track of the WDs evolution, while grayed out tracks show the pre-He-WDs evolution that goes through multiple Hydrogen flash cycles.
Red squares are the six previously known {\it Kepler} systems, green diamonds are the three previously known WASP systems, and blue circles are the four newly discovered systems. 
Gray dashed lines show isochrones of $0.2$, $0.4$, and $0.8$ Gyr, measured from the end of the mass-transfer epoch.
}
\label{fig:WDevo}
\end{figure*}

 Figure~\ref{fig:WDevo} shows the derived radii as a function of the $T_{\rm eff}$ of the newly discovered WDs, together with those of the previously known systems with derived radii, 
on top of WD evolution tracks
 of a set of WD masses from \citet{althaus13}.
Although \citet{althaus13} derived the tracks for WD orbiting a neutron star, they suggest that this evolutionary stage of the WD does not depend on the nature of its companion.
The figure shows that the known systems, together with KIC 4169521, all with hot and bloated WD secondaries, represent young systems probably at the pre-He-WD, or the initial WD cooling track stage \citep{vankerkwijk10,rappaport15}. The three new systems---KOI-3818, KIC 2851474, and KIC 9285587, are probably positioned further along the WD cooling track. 
Reading from the evolutionary tracks of Figure~\ref{fig:WDevo}, we estimate the ages
of the systems, measured from the end of the mass-transfer epoch, to be at the range of $0.2$--$0.8$ Gyr. These are roughly consistent with the ages of the A-star primaries that we derive from the stellar evolutionary tracks (see Figure~\ref{fig:evo} and Table~\ref{table:sys}).

The KOI-3818 system, with its fast-rotating primary and very small eccentricity, can shed some light on the dynamical evolution of the binary. 
On one hand, the fast rotation of KOI-3818 might indicate that the synchronization timescale is much longer than the age of the binary, measured from the end of the mass-transfer phase. On the other hand, tidal evolutionary models of short-period binaries predict synchronization and alignment timescales to be two to three orders of magnitude shorter than the circularization timescale \citep{zahn89,witte02}. We can therefore conclude that the present small eccentricity is not a product of tidal circularization at the present phase, but instead is the result of the mass-transfer process.
Evolutionary models of the mass-transfer phase do predict fast-rotating primaries and very small eccentricities \citep[e.g.,][]{toonen14}.

 The minute eccentricity component of $e \cos \omega=-0.00146(33)$ that we derive for KOI-3818 is about half of the eccentricity component of $e \cos \omega=0.0029(5)$ derived by \cite{carter11} for KIC 10657664.
Such small eccentricities are impossible to obtain from RV measurements alone, due to the small line-profile distortion expected for short-period binaries \citep[e.g.,][]{lucy05, komo07}. 
Small eccentricities have been measured previously for pulsars,\footnote{http://www.atnf.csiro.au/people/pulsar/psrcat/}
with WD secondaries in particular \citep[e.g.,][]{manchester05}, utilizing the precise timing of the observed radio pulses.  Obviously, evolutionary models \citep{toonen14} that include tidal interaction \citep[e.g.,][]{anton14} should account for these non-zero eccentricities.

It is also interesting to note that out of the four systems, only the primary of KOI-3818 is a fast rotator, while the other three derived rotation periods of the A-stars  are consistent with being synchronized. This may be because the binary separation to primary radius ratio, $(a/R_1)$, of KOI-3818 is relatively large, $6.9$, while it is much smaller for the other three binaries --- $2.7$, $3.7$, and $3.8$ for KIC 4169521, KIC 2851474, and KIC 9285587, respectively. As we can estimate the age of the four systems {\it from their WD radii and temperatures} (Figure \ref{fig:WDevo}), we might be able to constrain the calibration of the   
 synchronization timescale for stars with radiative envelopes. 

The exquisite {\it Kepler} photometry led to the discovery of 10 short-period dA+WD EBs, and has enabled to derive the masses, radii, and effective temperatures of both components of each system. Together with the 18 similar WASP systems these open a unique window into the end products of binaries.
 In particular, we have discovered three dA+WD systems with small,  $\lesssim 0.04$ $R_{\odot}$, WDs, extending the known population to older systems with cooler and smaller WD secondaries.

Finally, the identified sample of 10 short-period dA+WD eclipsing binaries in the {\it Kepler} stars should
enable us to estimate the statistics of this population and the fraction of A-stars with compact companions. The BEER search might help us to detect such non-eclipsing binaries lurking in the {\it Kepler} light curves.

\acknowledgments
We are thankful to the anonymous referee for comprehensive and valuable remarks that significantly improved this manuscript.
We thank Donald W. Kurtz for his review and advice regarding the high-frequency photometric modulations that we report here.
We are indebted to Andrew H. Szentgyorgyi, who led the TRES project, and to
Gabor F\H{u}r\'{e}sz for his many contributions to the success of the
instrument. We thank Perry Berlind, Gilbert A. Esquerdo, and Michael L. Calkins for obtaining the TRES observations, and
Allyson Bieryla and Jessica Mink for help with the data analysis.
We feel deeply indebted to the team of the Kepler mission, who enabled us to search and analyze their unprecedentedly accurate photometric data.
The research leading to these results has received funding from the European Research Council under the EU's Seventh Framework Programme (FP7/(2007-2013)/ ERC Grant Agreement No.~291352).
This research was supported by the ISRAEL SCIENCE FOUNDATION (grant No.~1423/11) and the Israeli Centers Of Research Excellence (I-CORE, grant No.~1829/12).
All of the photometric data presented in this paper were obtained from the Mikulski Archive for Space Telescopes (MAST). 
STScI is operated by the Association of Universities for Research in Astronomy, Inc., under NASA contract NAS5-26555. Support for MAST for non-HST data is provided by the NASA Office of Space Science via grant NNX09AF08G and by other grants and contracts.
We thank the Kepler mission for partial support of the spectroscopic
observations under NASA Cooperative Agreement NNX13AB58A with the
Smithsonian Astrophysical Observatory, DWL PI.

{\it Facility:} 
\facility{FLWO:1.5 m(TRES)}

\end{document}